\documentclass[lettersize,journal]{IEEEtran}

\usepackage{color,array,amsthm}
\usepackage{hyperref}
\usepackage{graphicx}
\usepackage{pifont}
\usepackage{url}
\usepackage{subfig}
\usepackage{booktabs}
\usepackage{fontawesome}
\usepackage{amsmath}
\usepackage{amssymb}
\usepackage{diagbox}

\usepackage{multicol} 
\usepackage{multirow} 
\usepackage[ruled]{algorithm2e}
\usepackage{bm}

\newcommand{\udm}[1]{\textcolor{black}{#1}}

\newcommand\Sec{Section.}
\newcommand\fig{Fig.}
\newcommand\tab{Tab.}
\newcommand\alg{Algorithm.}
\newcommand\eqt{Eq.}

\newcommand\sml{SML}

\newcommand\app{Appendix.}
\newcommand\pro{ReLMD}
\newcommand\solver{PLAMS}
\newcommand\sn{MCN}

\begin{document}
\title{Co-optimizing Topology and Geometry in Mega-Constellation Networks via Structural Motif-Lattice Paradigm}

\author{Xiangtong Wang\thanks{e-mail: wxton@foxmail.com}}

\maketitle
\begin{abstract}
Mega-constellation networks (MCNs) are currently revolutionizing global internet accessibility by providing ubiquitous connectivity on a planetary scale. 
  While the architectural configuration of MCNs is paramount to achieving high-performance space-based networking, the design process is inherently complex. 
  This complexity stems from the vast system scale and tightly coupled parameters, which culminate in a high-dimensional combinatorial optimization challenge. To address this challenge, we propose the Structural Motif–Lattice (SML) paradigm, a framework that decouples the MCN design space into two independent dimensions: topological connectivity (defining inter-satellite link logic) and geometric layout (defining the spatial distribution of satellites). 
  This decomposition is theoretically justified by the inherent separability of connectivity logic and spatial distribution in MCN architecture, thereby reducing the original high-dimensional problem to a tractable bi-dimensional optimization task. Within the SML paradigm, we formalize the Reliable and Low-latency MCN Design problem and develop the Progressive Motif and Lattice Search (\solver)  algorithm to find near-optimal MCN configurations. 
  
  Experiments conducted on major constellations including Starlink, OneWeb, Kuiper, and Telesat demonstrate that \solver~ achieves performance comparable to or better than state-of-the-art methods, yielding substantially enhanced network reliability and significant reductions in average propagation latency.

    \end{abstract}

   
\IEEEpeerreviewmaketitle

\section{Introduction}
Mega-constellation networks (MCNs), comprising thousands of low Earth orbit (LEO) satellites have emerged as a promising approach to provide global Internet access in recent years~\cite{handley2018delay,bhattacherjee2018gearing}. 
Commercial companies such as SpaceX~\cite{starlink}, Amazon~\cite{kuiper}, OneWeb~\cite{oneweb}, and Telesat~\cite{telesat} are actively competing in this `NewSpace' race. 
When integrated with terrestrial networks and equipped with inter-satellite links (ISLs), these systems offer the potential for worldwide connectivity with low latency and high bandwidth, moving beyond the limitations of traditional 
bent-pipe architectures.

The architectural design of a MCN is of critical importance, as it determines the upper bound of the network's operational capabilities.
However, such design must simultaneously account for the multidimensional interplay of constellation geometry, ISL topology, ground‑coverage and other related factors, all of which collectively introduce substantial design complexity.
Recent studies have attempted to mitigate the design complexity of MCNs through connection pattern constraints \cite{werner2001topological,wood2001internetworking}, heuristic search\cite{basak2023exploring,lai2024your}, or parameter dimensionality reduction\cite{basak2025leocraft}.
However, these approaches largely simplify the problem by operating at a single structural level, adopting a decoupled design paradigm that either optimizes ISL connectivity under a fixed constellation geometry \cite{bhattacherjee2019network,han2024non,guo2024constellation} or adjusts geometric parameters according to predefined connectivity rules \cite{basak2025leocraft,zhang2024depth,lan2024inter}. 
While such decoupling reduces the search space, it imposes strong structural assumptions that exclude potentially high‑performing configurations and hinder structural exploration of the joint design space. 

From the perspective of structural modeling, the aforementioned challenges stem from a lack of higher-level abstraction and decoupled understanding of MCN structural elements. 
To address this, we propose a key insight: the overall structure of an MCN is fundamentally governed by two independent and theoretically separable components. 
The first is \textit{topological connectivity}, which determines the inter‑satellite connectivity. 
The second is \textit{geometric layout}, which dictates the spacial arrangement of satellites within the constellation.
Crucially, once these two high-level structural aspects are determined, numerous lower-level parameters (e.g., number of orbits, phase factor) can be optimized jointly, thereby dramatically simplifying the design space. 
This independence is validated by the fact that connectivity patterns (e.g., +Grid\cite{bhattacherjee2019network}, triangular\cite{Kedrowitsch2024ResilientRF}) can be instantiated across different geometric arrangements (e.g., rectangular, hexagonal lattices) without mutual interference.

Building on this structural insight, we therefore propose the \textbf{S}tructural \textbf{M}otif-\textbf{L}attice (\textbf{SML}) paradigm for MCN design. This framework explicitly decouples the MCN design problem into two corresponding elements: \textit{motif}, which defines the topological connectivity of the network, and \textit{lattice}, which governs the geometric layout of satellites.
\begin{figure}[t!]
    \begin{center}
        \includegraphics[width=1\linewidth]{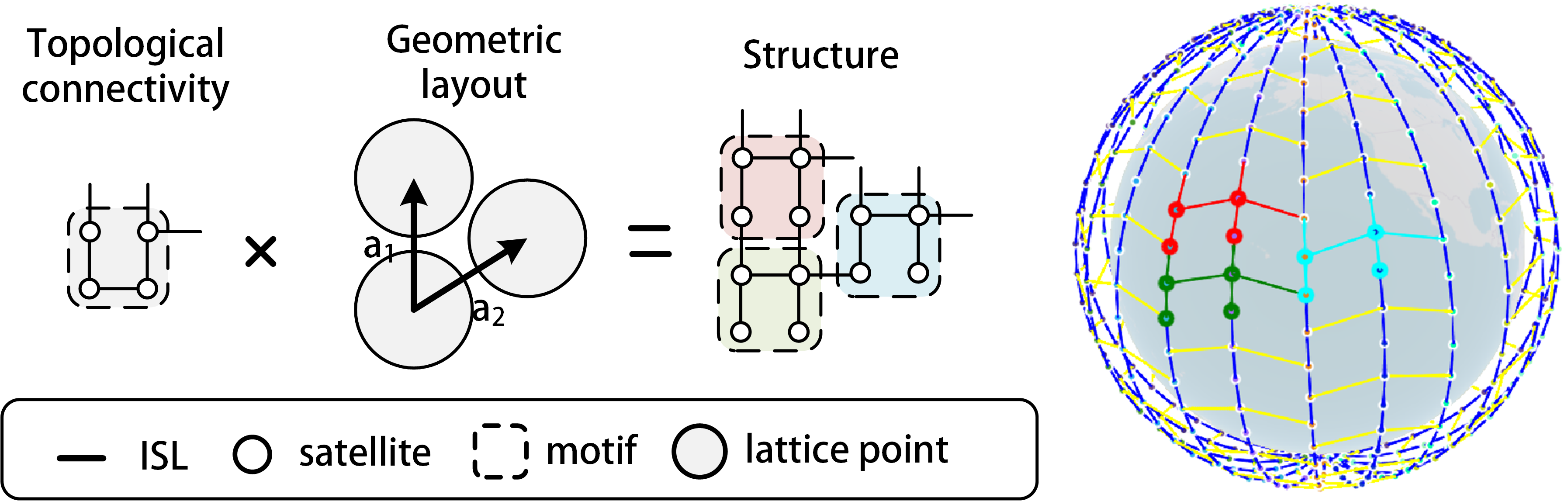}
    \end{center}
    \vspace{-1em}
    \caption{The two‑dimensional SML representation: motifs define inter‑satellite connectivity, while lattices govern the periodic spatial arrangement of motifs. 
    The primitive vectors $\mathbf{a}_1$ and $\mathbf{a}_2$ determine the lattice geometry and replication rules, as detailed in \Sec\ref{sec:lattice}.}
    \vspace{-1em}
        \label{fig:teaser}
 \end{figure}
 \fig\ref{fig:teaser} illustrates the core concept of the SML paradigm. The SML paradigm constructs an MCN by encoding inter‑satellite connectivity through motif, and replicating this motif across the constellation according to a predefined lattice. This two‑dimensional decomposition separates the design space into independent aspects of topological connectivity and geometric arrangement, thereby enabling structural and efficient MCN designing pipeline.

By separating connectivity and geometry, this abstraction reduces the original high-dimensional design problem to an optimization over a compact, low-dimensional space, while preserving physical feasibility and the expressive capacity for diverse topologies. 
Within this paradigm, designers can independently explore geometric configurations (e.g., satellite arrangement, orbital phase factor) and topological motifs (e.g., +Grid, xGrid, or hybrid), enabling a structural and scalable search across heterogeneous MCN architectures.

Beyond establishing the SML design paradigm, a key question is how to configure an MCN for optimal performance. 
While high-level metrics such as routing efficiency, network capacity, and operational cost are commonly considered\cite{werner2001topological,guo2024constellation,rabjerg2021exploiting}, they ultimately depend on two fundamental structural properties determined by the motif and lattice: network reliability, which measures the ability to maintain connectivity and sustain traffic, and end-to-end latency, which reflects the traffic transmission efficiency. 
These properties underpin all higher-level metrics and exhibit an inherent trade-off, posing a central challenge in MCN design. 
\udm{From a configuration optimization perspective, reliability and latency naturally serve as objective metrics. }
To address this, we formulate the Reliable and Low-latency MCN Design (\textbf{ReLMD}) problem, which leverages the SML paradigm to identify motif–lattice configurations that balance these competing objectives under realistic operational constraints, enabling high-available and efficient network architectures.

In summary, the main contributions of this paper are as follows:

\noindent\ding{108} We propose SML (Structural Motif-Lattice), a new MCN design paradigm that decouples MCN structure into topological connectivity and geometric layout, reducing the design space from high-dimensional to two independent dimensions.

\noindent\ding{108} By leveraging SML paradigm, we formalize the High-reliability and Low Latency MCN Design (\pro) problem, aiming to maximum the reliability of MCN while minimizing the traffic latency.

\noindent\ding{108} We develop Progressive Lattice And Motif Search (\solver) algorithm and apply it to several operational mega-constellations for their optimization. Based on a comprehensive evaluation, the results show that the optimized network architectures achieve either superior or competitive performance compared to the original configurations.

The remainder of this paper is organized as follows: 
\Sec\ref{sec:bg} gives some literature review of our work.
\Sec\ref{sec:sml} presents the Structural Motif-Lattice MCN construction paradigm.
In \Sec\ref{sec:pro}, we formulate the \pro~ problem and propose a heuristic algorithm \solver~ to solve it.
We optimize the design of the existing MCN system in \Sec\ref{sec:exp} based on SML paradigm.
Finally, \Sec\ref{sec:limitations} and \Sec\ref{sec:con} discuss limitations and conclude the paper.

\section{Related work}
\label{sec:bg}

Research on MCN design has evolved along a specific trajectory, progressing from 
constellation geometry design, 
topology planning, topology design, constellation parameterized search and 
sub-structure based design.

\subsection{Constellation Geometry Design}
\label{sec:related1} 
Early constellations with limited number of satellites were primarily designed to maximize coverage of a target area using as few satellites as possible, while minimizing collision risk or operational cost (satellite count). Two dominant configurations emerged during this era: the Walker constellation~\cite{walker1984satellite,ballard1980rosette}, which provides uniform global coverage, and the street-of-coverage configuration~\cite{luders1961satellite,rider1986analytic}, tailored for continuous coverage over specific zones.
More recent works~\cite{deng2021ultra} have extended this paradigm to MCN design by incorporating non-uniform terrestrial traffic demands, aiming to reduce satellite count while meeting user-specific service requirements.
Nevertheless, these studies largely center on the trade-off between geometric layout and ground traffic distribution, with limited attention to the networking performance of satellite networks enabled by ISLs.

\subsection{Topology Planning}
\label{sec:related2} 

Recent research has paid attention to enhance the networking capabilities of satellite constellations through performance-aware topology planning, taking into account routing efficiency~\cite{chang1995topological,chang1998fsa,Kedrowitsch2024ResilientRF,werner2001topological,rabjerg2021exploiting,han2024non}, network capacity~\cite{guo2024constellation,lan2024inter,leyva2021inter}, throughput \cite{higashimori2024dynamic} and energy consumption~\cite{alagoz2011energy,leyva2021inter,zhou2017mission}.
More specifically, these works can be categorized from two dimensions in terms of ISL planning:
(i) ISL technology, distinguishing between optical and RF ISLs, where optical ISLs often impose stricter limits on the number of maintaining ISLs per satellite; and
(ii) ISL persistence, differentiating temporary from permanent ISLs, where temporary ISLs require frequent inter-satellite handovers, whereas permanent ISLs maintain a fixed connection pattern over time.
Many above approaches treat ISLs as resources and employ planning algorithms to dynamically deactivate certain ISLs or satellites, thereby reducing operational costs while maintaining acceptable performance.
However, highly dynamic, performance-aware ISL reconfiguration demands a centralized control plane, which poses significant scalability challenges in future mega-constellations.

\subsection{Topology Design}
\label{sec:related3} 

In contrast to topology planning, topology design in satellite networks focus on specific connection patterns and analyzing their impact on the network. 
This includes examining dynamic characteristics between satellites \cite{suzuki2007study,wang2007topological} as well as overall network performance. 
Early work by \cite{wood2001internetworking} provided a extensive analysis of the network performance for basic topologies, and much of the subsequent research has concentrated on ``+Grid'' topologies \cite{basak2023exploring,guo2024constellation,lan2024inter,leyva2021inter,lai2024your,zhang2024depth}, which denotes 4-ISL connection pattern with nearest neighbors along orbit and phase directions. However, many other topological types exist beyond ``+Grid''. 
For example, \cite{suzuki2007study} highlighted the superior north-south connectivity offered by the figure-of-eight ISLs in the GlobalStar project \cite{dietrich1998globalstar}. 
Leveraging information from SpaceX, \cite{chen20243} simulated a 3-ISL structure with lower link maintenance costs and evaluated its routing performance. 
Meanwhile, \cite{mclaughlin2023grid} proposed the ``xGrid'' model, which essentially constitutes a specialized form of inter-plane ISL between more distant satellites. 
Additionally, \cite{Kedrowitsch2024ResilientRF} investigated resilient routing within triangular topological patterns. 
However, most existing studies evaluate topology performance only under fixed constellation parameters. This limitation precludes an understanding of how variations in orbital altitude, inclination, or satellite count interact with topological behavior, thereby constraining systematic performance analysis across the design space.

\subsection{Parameterized Search in Constellation Design}
\label{sec:related4} 

As MCNs scale up, optimization-based design approaches have been increasingly proposed to explore the vast design space of MCNs.
Several studies~\cite{basak2023exploring,lai2024your,zhang2024depth} formalize the MCN design problem as a parameter search under given constraints, aiming to identify optimal constellation configurations (e.g., number of planes $N_p$, number of satellites per plane $M_p$, and phase factor $F$). However, the high dimensionality of the search space renders exhaustive exploration intractable, and these works largely restrict topology to the conventional ``+Grid'' pattern, omitting systematic topological optimization.
More recently, Basak et al.~\cite{basak2025leocraft} introduced a flexible and efficient simulation framework that reduces optimization complexity by grouping design parameters. Their work evaluates several connectivity patterns beyond ``+Grid'' and provides a comprehensive performance comparison. 
Nevertheless, topological choices remain coupled to orbital parameters $(N_p, M_p, F)$, rather than being treated as independent design variables within the search space. 
This coupling inherently restricts the representation of more sophisticated and structural topologies, such as the 3‑ISL configuration\cite{chen20243}, potentially excluding high‑performing design alternatives. 
This limitation stems largely from the inefficiency of existing topology representation methods: encoding connectivity through adjacency matrices would entail a $2^N$ search space ($N$ is number of ISLs), which is computationally infeasible and severely restricts systematic topology exploration.

\subsection{Sub-structure based Design Paradigm}
\label{sec:related5} 

MCNs exhibit a pronounced periodic structure in their spatial and topological organization. This characteristic aligns with insights from complex network theory, which suggests that large-scale networks can be effectively represented as assemblies of recurring small-scale substructures \cite{milo2002network}. Such a substructure-based perspective provides a higher-level abstraction for topology representation, thereby enabling more efficient and scalable topology optimization.
Inspired by this, recent works~\cite{bhattacherjee2019network,zhao2022self,han2024non,lin2022inter} propose designing MCNs based on motifs, offering a novel paradigm for MCN topology design where ISLs are not restricted to nearest-neighbor connections and can support diverse connection patterns. 
While existing work primarily leverages motifs to enhance routing resilience \cite{zhao2022self} or latency\cite{bhattacherjee2019network}, we contend that employing a simple, universal substructure to represent the entire network topology holds significant potential for enabling efficient and flexible topology search.

In summary, prior research on MCN design faces a fundamental tension between scalability and expressiveness: it either addresses high‑dimensional orbital optimization at the expense of topological diversity\cite{basak2023exploring,zhang2024depth,lai2024your,basak2025leocraft}, or explores richer topologies while remaining constrained by fixed orbital parameters \cite{suzuki2007study,chen20243,zhao2022self}. This limitation motivates our pursuit of an efficient design paradigm that jointly optimizes topology and geometry within a compact, low‑dimensional search space.

\begin{figure*}[t!]
    \centering
    \includegraphics[width=.75\linewidth]{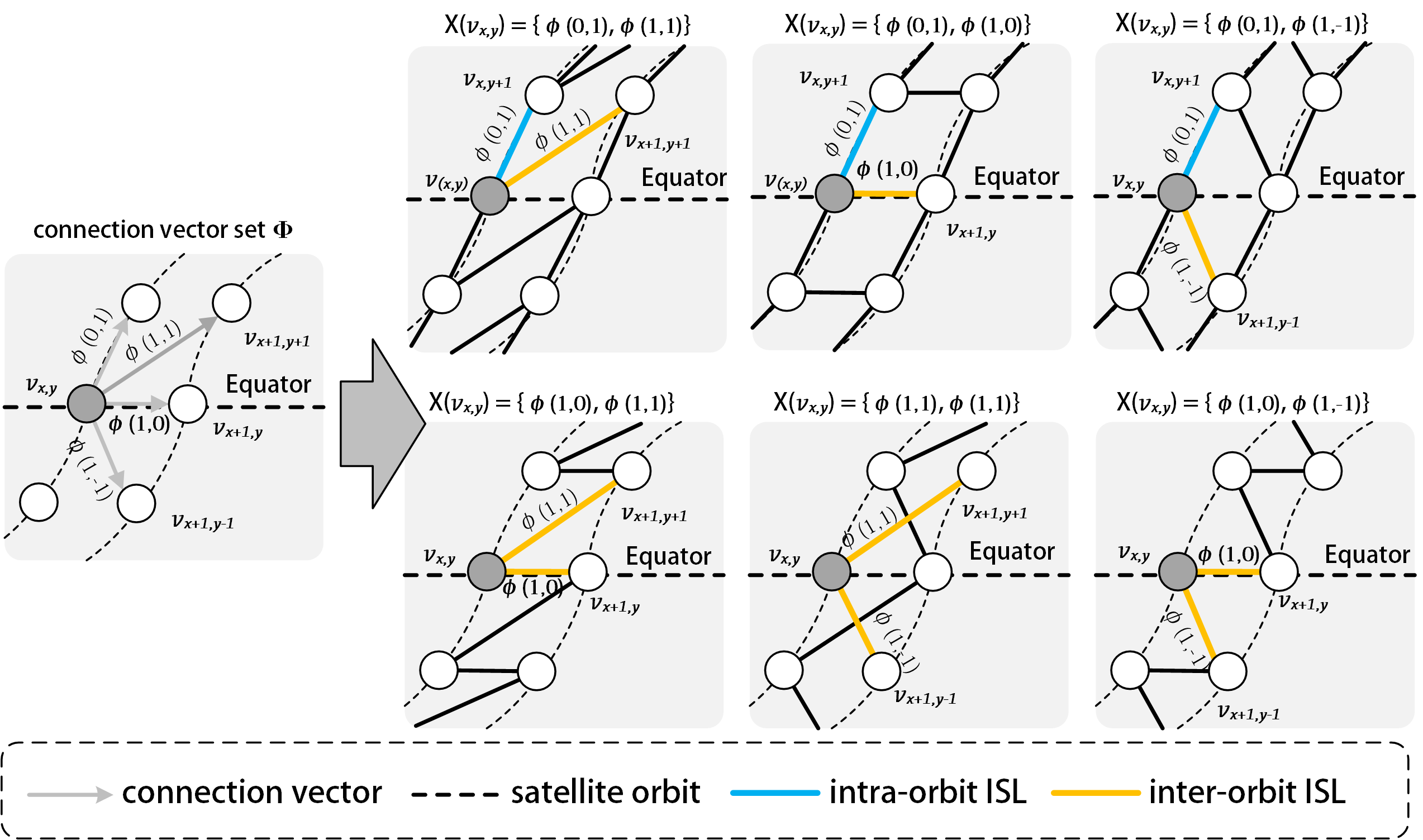}
    \caption{Connection vectors and their composed connection patterns.}
    \label{fig:cf}
\end{figure*}

\section{Structural Motif--Lattice Paradigm for MCN Design}
\label{sec:sml}

\subsection{Walker-Based Mega-Constellation Network Model}
\label{sec:mcn}
We base our work on the widely used Walker constellation model before introducing the SML paradigm. Walker constellations are widely adopted in operational MCNs (e.g., Starlink\cite{starlink}, OneWeb\cite{oneweb}) due to their uniform global coverage, configurable parameters, and periodic, grid-like structure, which naturally supports decomposition into orthogonal topological and geometric dimensions. We first formalize the Walker-based MCN model to provide the foundation for subsequent decomposition into motif and lattice.

\subsubsection{Walker constellation}

A Walker constellation~\cite{walker1984satellite} provides uniform global coverage and is parameterized by the notation~$N_p / M_p / F / \theta$, where $N_p$ is the number of orbital planes, $M_p$ is the number of satellites per plane, $F \in \{0, 1, \dots, N_p-1\}$ is the phase factor, and $\theta$ is the orbital inclination.
In a Walker constellation, the phase differences between adjacent satellites in the same plane ($\Delta f$) and in neighboring planes ($f_0$) are respectively given by:
\begin{equation}
\Delta f = \frac{2 \pi }{M_p},\quad
 f_0 = \frac{2 \pi F}{N_p \cdot M_p}
\end{equation}
The right ascension of the ascending node (RAAN) spacing~$\Delta \Omega$ between adjacent planes depends on the constellation type. Two principal variants exist: Walker-$\delta$ and Walker-Star. The Walker-$\delta$ configuration, with $\Delta \Omega = \dfrac{2\pi}{N_p}$, is typical of inclined-orbit constellations. The Walker-Star configuration, with $\Delta \Omega = \dfrac{\pi}{N_p}$, predominantly employs polar or near-polar orbits.
Note that the constellation configurations are identical when $F = 0$ and $F =  N_p$.
The maximum phase offset, $\Delta f = \pi / M_p$, occurs when $F = \lfloor N_p / 2 \rfloor$, where $\lfloor \cdot \rfloor$ denotes the floor operator.

\subsubsection{Mega-constellation network}  
Modern mega-constellations extend the basic Walker model by incorporating inter-satellite links (ISLs), enabling packet forwarding between satellites and forming a mega-constellation network (MCN). The resulting MCN structure $\mathcal{S}$ can be formally represented as $\mathcal{S}(N_p,M_p,F,\theta,A)$, where $N_p,M_p,F,\theta$ denote the constellation parameters
 , while $A$ represents the adjacency matrix (a binary matrix where $A_{i,j}=1$ if satellite $i$ and satellite $j$ are connected by an ISL, and 0 otherwise). 
 The inclusion of constellation parameters and topological connectivity $A$ significantly expand the design space, rendering MCN structure optimization a high-dimensional combinatorial problem.

\subsection{Decomposing MCN Structure into Motif and Lattice}
MCNs constructed on Walker configurations exhibit strong structural periodicity. In materials science, periodic structures are compactly described by the repetition of a fundamental unit within a lattice framework. Inspired by this principle, we propose a novel MCN design paradigm, Structural Motif-Lattice (SML), in which the network structure is represented as the combination of a motif and a lattice.

Under the SML paradigm, MCN design is decomposed into two independent components: the motif $\mathcal{M}$, which characterizes local topological connectivity, and the lattice $\mathcal{L}$, which specifies the global geometric layout. Formally, an MCN structure is expressed as:
\begin{eqnarray}
    \mathcal{S}(\mathcal{M}, \mathcal{L})
\end{eqnarray} 
 where the motif $\mathcal{M}$ is defined through a hierarchical abstraction of satellite nodes, connection vectors, and connection patterns, providing a Structural representation of logical connectivity. The lattice $\mathcal{L}$ adopts crystallographic models to describe the periodic spatial arrangement of motifs, thereby determining the constellation geometry.
This decomposition is theoretically justified by the separability between connectivity logic and spatial distribution: local connection rules are independent of the global satellite placement, enabling independent exploration and optimization of topology and geometry within a unified design framework.

\subsection{Topological Connectivity Design Using Motifs}
\label{sec:motif}

To formalize the concept of motif, we leverage the grid-like Walker structure to establish a unified coordinate system, which serves as the foundation for defining connection vectors. These vectors are aggregated into connection patterns for individual satellites, which ultimately constitute the motifs. This hierarchical framework enables a compact and structural representation of the constellation's topological connectivity.

\subsubsection{Orbit–phase coordinate system (OPCS)} 
We define an Orbit–Phase Coordinate System (OPCS), in which each satellite is uniquely indexed by a pair of discrete coordinates:
\begin{eqnarray}
v_{x,y}, ~x \in [0, N_p-1], \; y \in [0, M_p-1],
\end{eqnarray}
where $x$ denotes the orbital plane index and $y$ denotes the phasing index within the plane. 

\subsubsection{Connection Vectors}
Given an ISL between satellites $v_{x_i,y_i}$ and $v_{x_j,y_j}$, their relative connectivity in OPCS is represented by a connection vector:
\begin{eqnarray}
      \phi(\Delta x,\Delta y)
\end{eqnarray}
where
    \begin{align}
        \label{eqt:cv}
        \Delta x &= (x_j-x_i) \mathop{mod} N_p \\
        \Delta y &= \left\{
        \begin{aligned}
            (y_j-y_i)\mathop{mod} M_p, &\text{if}~x_i < N_p-1 \\
            (y_j-y_i-F) \mathop{mod} M_p, &\text{if}~x_i = N_p-1
        \end{aligned}
        \right.
    \end{align}
Here, $\mathop{mod}$ denotes the modulo operation. 
 When the connecting satellite lies in the last orbital plane ($x_i = N_p-1$)an additional subtraction of $F$ is required for $y$ to prevent geometric distortion.

\subsubsection{Connection patterns}
A satellite’s connection pattern is defined as the set of connection vectors it maintains, which fully characterizes its local ISL configuration. 
While a single \udm{connection vector} defines a relative ISL direction, the connection pattern specifies the entire set of neighbors to which a satellite is simultaneously connected. Let $\Phi$ denote the set of all feasible connection vectors in OPCS. 
Given a hardware constraint limiting each satellite to at most $c_{max}$ distinct vectors, the admissible connection patterns of satellite $v_{x,y}$ are defined as
\begin{eqnarray}
    X(v_{x,y}) \in \mathcal{X}= \bigcup\limits_{c=1}^{c_{\max}} \binom{\Phi}{c}
\end{eqnarray}
where $\binom{\Phi}{c}$ denotes the $c$-element subsets of $\Phi$.

\fig\ref{fig:cf} illustrates the possible connection vectors and its composed connection patterns for a satellite $v_{x,y}$ with $c=2$ connection vectors, representing both intra-plane (\textcolor{blue}{blue lines}) and inter-plane (\textcolor{yellow}{yellow lines}) connectivity. These compositions lead to distinct periodic network topologies in the MCN. 

\subsubsection{Motif}
Building upon the definitions of the OPCS, connection vectors and connection patterns, a motif is a structural local connectivity unit consisting of $d$ satellites, serving as a functional template for periodic replication. Formally, a motif $\mathcal{M}$ of size $d$ is represented by the tuple:
\begin{eqnarray}
    \mathcal{M}(\mathbf{R},\mathbf{X})
\end{eqnarray}
where, $\mathbf{R}= (v_1,\cdots,v_d)$ defines the relative spatial arrangement of satellites in OPCS; 
$\mathbf{X}=(X_1,\cdots, X_d)$ assigns a connection pattern $X_i\in\mathbf{X}$ to each satellite $v_i$.
The ordered correspondence between $\mathbf{R}$ and $\mathbf{X}$ uniquely determines the motif’s topological structure. By varying the motif size $d$, one can represent single-satellite motifs, multi-satellite motifs, or larger composite structures.

\begin{figure}[t!]
    \centering
    \includegraphics[width=0.85\linewidth]{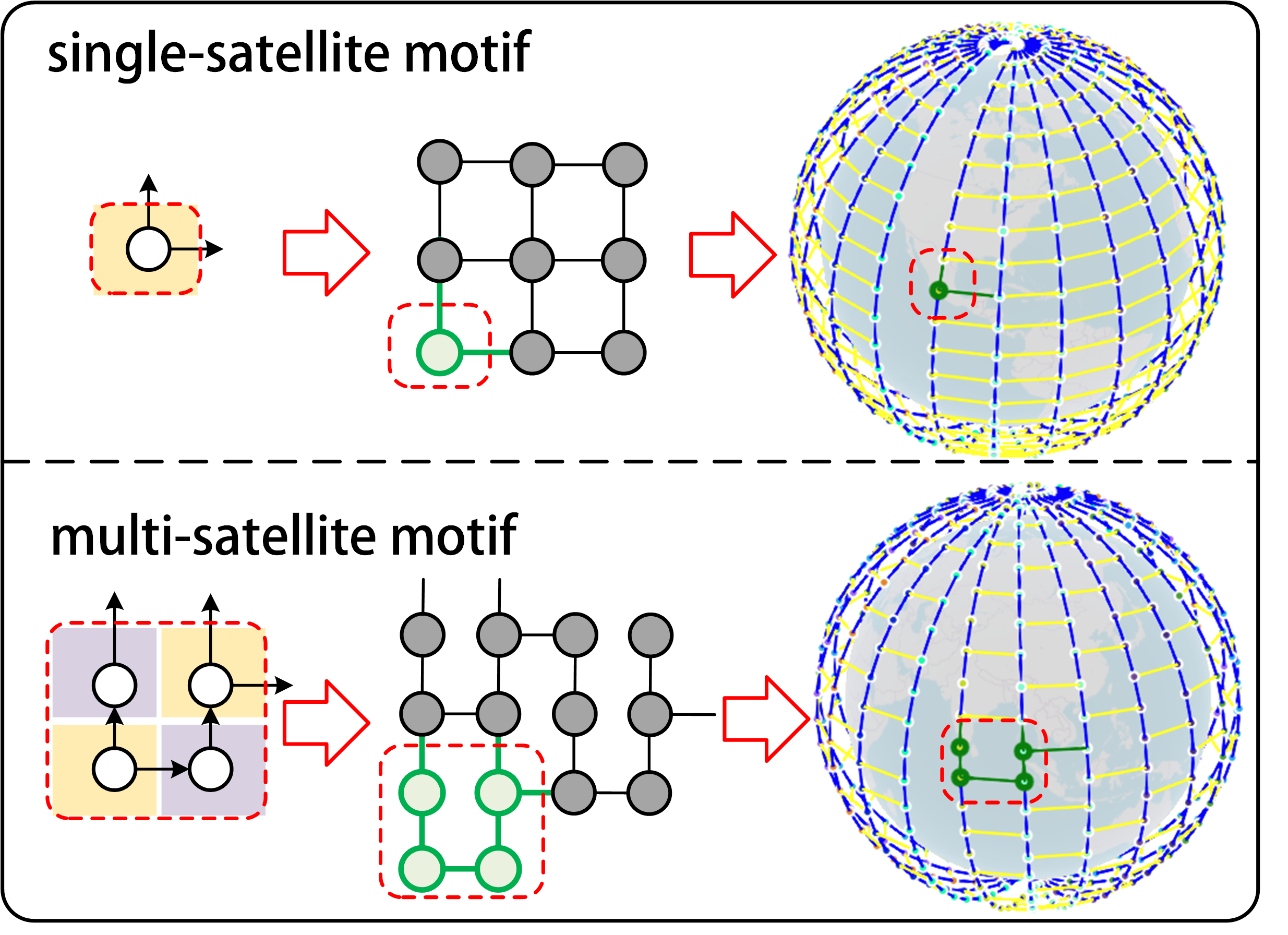}
    \caption{
        The MCN structure is built from motifs, each of which is defined by spacial arranged connection patterns.
    }
    \label{fig:motif}
\end{figure}

\fig\ref{fig:motif} illustrates examples of single‑satellite and multi‑satellite motifs, each characterized by a distinct connection pattern (represented by colored blocks) arranged in a particular spatial arrangement. 
By designing increasingly complex motifs, the overall network topology can be systematically organized to follow predefined sub-structure cycles.

\subsection{Geometric Layout Design Using Lattices}
\label{sec:lattice}

After completing the design of local topological connectivity, it is also necessary to specify how these local units are repeatedly arranged on a global scale. 
The Walker constellation possesses strict structural periodicity: orbital planes are uniformly spaced in RAAN at equal angular intervals of $360^\circ/N_P$; satellites within each plane are equally spaced in phase at intervals of $360^\circ/M_P$; and the phase offset between satellites in adjacent orbital planes is fixed by the phase factor $F$, being a fraction of the intra-plane spacing. 
 This construction yields a set of uniformly distributed discrete points in the two-dimensional projection plane, where the relative positional relationship between each satellite and its neighbors remains consistent throughout the constellation. This closely mirrors the essential meaning of a ``lattice'' in crystallography: a set of isotropically repeating discrete grid points generated by the translation operations of a finite number of basis vectors. To this end, this chapter adopts the lattice concept from crystallography to describe the global geometric layout of an MCN, referring to it as the lattice.


\begin{figure}[t!]
    \centering
    \includegraphics[width=1\linewidth]{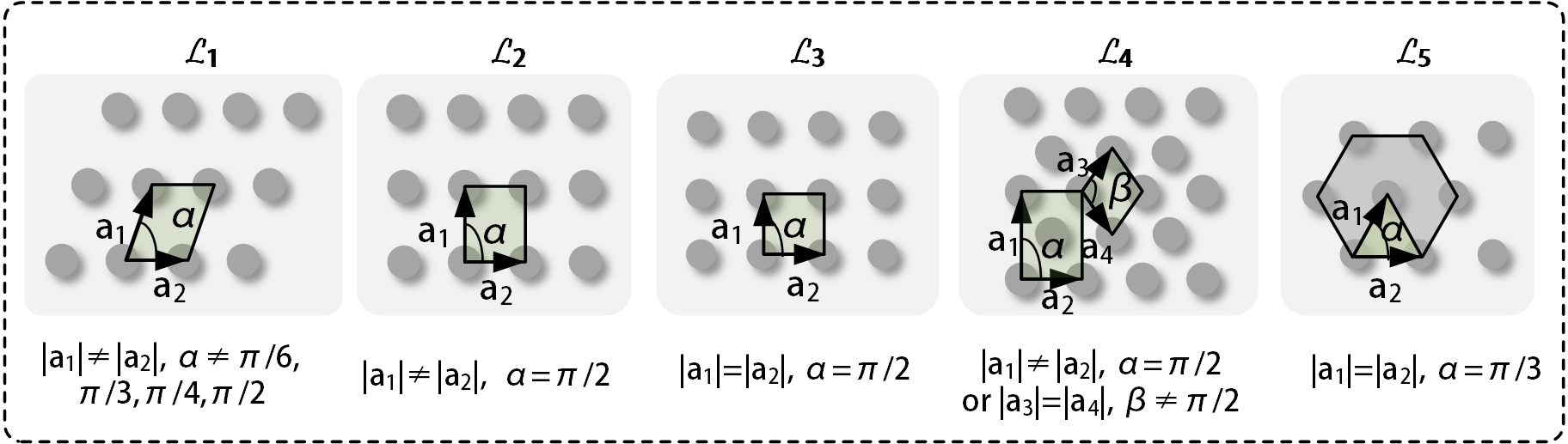}
    \caption{Five possible MCN geometric layouts based on Bravais lattices theory \cite{campbell1986historical}. The black dash line presents the Earth Equator while the grey dash line stands for the satellite orbit plane.}
    \label{fig:lattice}
\end{figure}

Bravais lattice theory~\cite{campbell1986historical} states that in two-dimensional space, all point arrangements capable of periodically and seamlessly tiling the plane belong to one of five fundamental types, as shown in \fig\ref{fig:lattice}. Here, the gray points represent lattice points, corresponding to the positions of motifs, while $\mathbf{a}_1$ and~$\mathbf{a}_2$ denote the primitive translation vectors characterizing the lattice~\cite{Ashcroft1976}; their detailed geometric definition is provided in \app\ref{app:pvec}. 
Consequently, for a Walker constellation exhibiting periodic tiling behavior under two-dimensional projection, its geometric layout inherently falls into one of these five fundamental lattices.


\fig\ref{fig:walker-lattice} illustrates the five Bravais lattice types of Walker-$\delta$ MCN layouts under the single-satellite motif scenario (where the motif contains only one satellite). 
In this representation, the vector~$\mathbf{a}_1$ is typically aligned with the orbital direction, corresponding to the replication of motifs (satellites) along the same orbital plane; $\mathbf{a}_2$ characterizes the translation of satellites between adjacent orbital planes, thereby describing the arrangement of motifs (satellites) across orbital directions. Based on the relative magnitudes of~$|\mathbf{a}_1|$ and~$|\mathbf{a}_2|$ and their included angle $\alpha$, Walker constellation layouts can be classified into five representative lattice types. To mitigate the effects of polar projection distortion, the following classification is primarily based on the positional relationships of satellites near the equator:

\begin{figure*}[t!]
    \begin{center}
            \includegraphics[width=0.9\linewidth]{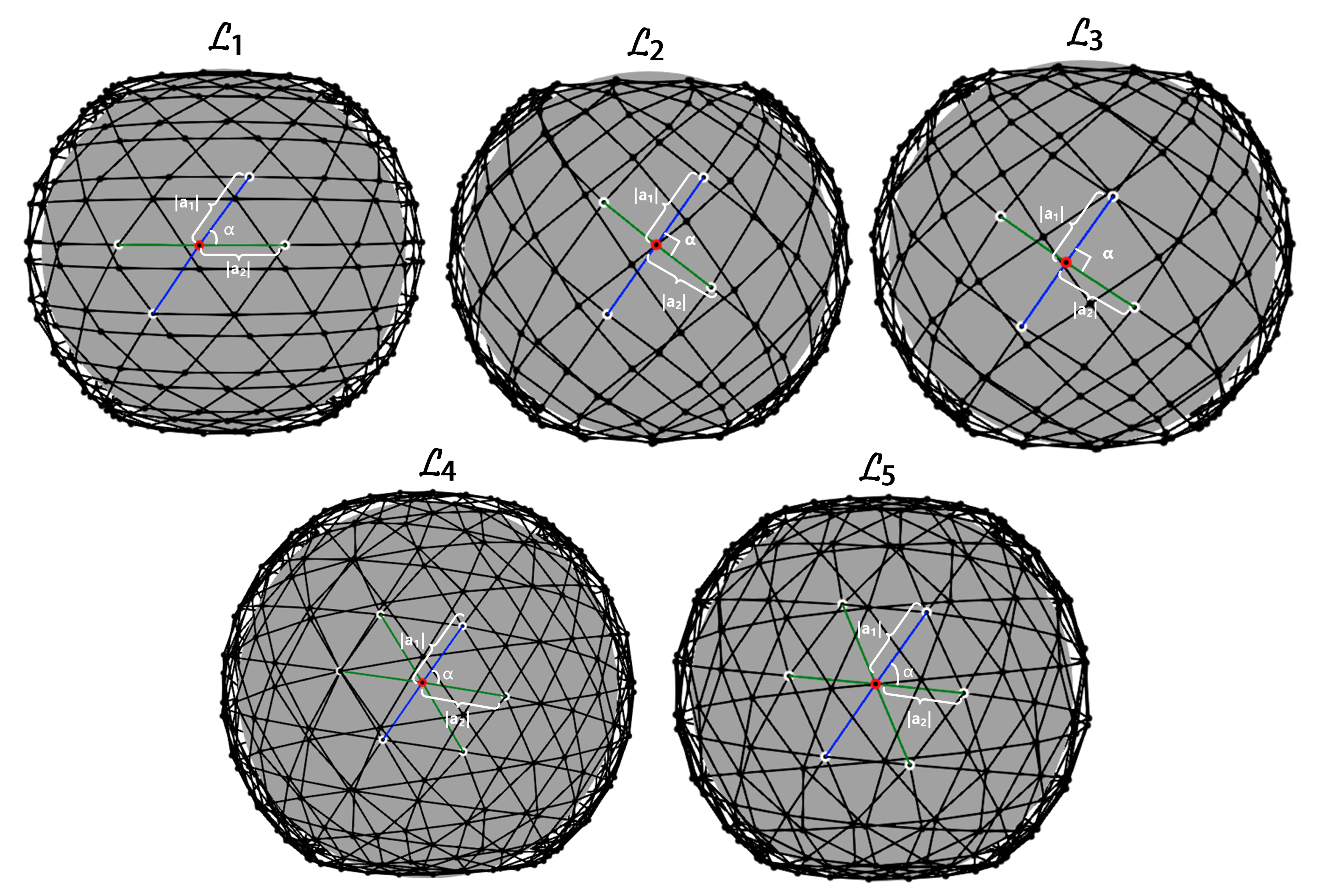}
    \end{center}
    \caption{
       The geometric layout of MCNs corresponding to five two-dimensional Bravais lattices under the single-satellite motif scenario.} 
        \label{fig:walker-lattice}
 \end{figure*}

\begin{itemize}


\item \textbf{$\mathcal{L}_1$ (Oblique layout)}:
The distances from a satellite to its nearest intra-plane and inter-plane neighbors are unequal, and the angle between the corresponding ISLs is neither~$30^\circ$, $60^\circ$, $45^\circ$, nor~$90^\circ$.


\item \textbf{$\mathcal{L}_2$ (Rectangular layout)}:
The distances from a satellite to its nearest intra-plane and inter-plane neighbors are unequal, and the angle between the corresponding ISLs is approximately $90^\circ$.


\item \textbf{$\mathcal{L}_3$ (Square layout)}:
The distances from a satellite to its nearest intra-plane and inter-plane neighbors are equal, and the angle between the corresponding ISLs is approximately $90^\circ$.


\item \textbf{$\mathcal{L}_4$ (Hexagonal layout)}:
The distances from a satellite to its nearest intra-plane and inter-plane neighbors are equal, but the angle between the corresponding ISLs is not~$60^\circ$.


\item \textbf{$\mathcal{L}_5$ (Regular hexagonal layout)}:
The distances from a satellite to its nearest intra-plane and inter-plane neighbors are equal, and the angle between the corresponding ISLs is~$60^\circ$.

\end{itemize}


By reconfiguring the number of orbital planes~$N_P$, the number of satellites per plane~$M_P$, and the phase factor~$F$, the relative lengths of the translational basis vectors and their included angle can be adjusted, thereby altering the lattice layout. This means that the global geometric structure of an MCN can be reshaped by reconfiguring constellation parameters without modifying the underlying local connectivity rules.

\subsection{Reconfiguring Geometric Parameters into Lattice Layouts}
\label{sec:lattice-adj}


The five lattice layouts described above impose specific geometric constraints on the constellation parameters ($N_p$, $M_p$, and $F$). However, the given original constellation parameters do not necessarily satisfy these constraints, and the layout typically corresponds to the most general~$\mathcal{L}_1$ lattice. To modify the constellation layout to a target lattice, it is necessary to derive reconfiguration parameters from the original constellation parameters that satisfy the corresponding geometric constraints. 
 
%

\tab\ref{tab:lattice} summarizes the constellation parameter mapping formulas corresponding to each lattice layout, where~$N_M$ is the total number of satellites in the original constellation and $\theta$ is the orbital inclination.
For the adjusted constellation, the number of orbital planes, the number of satellites per plane, and the phase factor are denoted by~$N_p^\star$, $M_p^\star$, and~$F^\star$, respectively. In addition, $f_0^\star$ denotes the phase difference between adjacent satellites in neighboring planes, $\Delta f^\star$ the phase difference between adjacent satellites within the same plane, and~$\Delta \Omega^\star$ the RAAN spacing between neighboring planes.

\begin{table*}[t!]
	\centering
	\caption{Parameter mapping for reconfiguring Walker constellations to target lattice layouts}
	\label{tab:lattice}
	\renewcommand{\arraystretch}{3} 
	\scalebox{1}{
		\begin{tabular}{cccc}
			\toprule[1.5pt]
			           & $N_p^\star$ & $M_p^\star$ & $F^\star$ \\
			\midrule
			$\mathcal{L}_2$ 
				& - 
				& - 
				& $
					\left\{
						\begin{aligned}
							\left\lvert N_p - M_p \cos (\theta) \right\rvert  & , \quad \text{Walker-}\delta \\
							\left\lvert N_p - \dfrac{M_p \cos (\theta)}{2} \right\rvert & ,\quad \text{Walker-star}
						\end{aligned}
					\right.
				$ \\
			$\mathcal{L}_3$ 
				& $
					\left\{
						\begin{aligned}
							& \sqrt{N_M \cos (\theta) \tan (\theta)} , \quad \text{Walker-}\delta \\
							& \sqrt{\dfrac{N_M \cos (\theta) \tan (\theta)}{2}} , \quad \text{Walker-star}
						\end{aligned}
					\right.
				$ 
				& $ \dfrac{N_M}{N^\star} $ 
				& $ \dfrac{(\tan (\theta) - 1) N_p^\star}{\tan (\theta)} $ \\
			$\mathcal{L}_4$ 
				& - 
				& - 
				& $
					\left\{
						\begin{aligned}
							\left\lvert \dfrac{N_p}{2} - M_p \cos (\theta) \right\rvert  & , \quad \text{Walker-}\delta \\
							\left\lvert \dfrac{N_p - M_p \cos (\theta)}{2} \right\rvert  & ,\quad \text{Walker-star}
						\end{aligned}
					\right.
				$ \\
			$\mathcal{L}_5$ 
				& $
					\left\{
						\begin{aligned}
							\sqrt{\dfrac{2 N_M \sin (\theta)}{\sqrt{3}}} , \quad \text{Walker-}\delta \\
							\sqrt{\dfrac{N_M \sin (\theta)}{\sqrt{3}}} ,\quad \text{Walker-star}
						\end{aligned}
					\right.
				$ 
				& $ \dfrac{N_M}{N^\star} $ 
				& $ \dfrac{ \sin (\pi/3 - \theta) N_p^\star }{ \sin (\theta) } $ \\
			\toprule[1.5pt]
		\end{tabular}
	}\\
	\footnotesize All~$F^\star$, $N_p^\star$, and~$M_p^\star$ are rounded to the nearest integer.
	\normalsize  
\end{table*}

In practice, $M_p^\star$, $N_p^\star$, and $F^\star$ should all be rounded to the nearest integer. Through the above reconfiguration, the original Walker constellation can be adapted to meet the parameter requirements of different lattice layouts, thereby forming constellation networks with diverse geometric structures. The detailed derivation of the parameter mapping formulas in \tab\ref{tab:lattice} is presented in \app\ref{app:reconfig}.

\subsection{Performance Advantages of Lattice Layouts}

Different lattice layouts have a significant impact on the geometric structure and ISL connectivity of constellation networks. The~$\mathcal{L}_1$ lattice is the common parallelogram layout, while~$\mathcal{L}_2$ and~$\mathcal{L}_3$ are optimized to rectangular and square connections, respectively, enhancing the connectivity of traffic paths in the direction perpendicular to the orbital plane, so that paths close to this direction have fewer zigzags and hops. Similarly, $\mathcal{L}_4$ and~$\mathcal{L}_5$ further improve multi-directional connectivity and ISL uniformity through triangular and hexagonal structures. Under these layouts, network ISLs are more evenly distributed in all directions and ISL lengths are more consistent, providing more equal-cost path choices for routing and thereby effectively reducing the fluctuation of delay and hop count with traffic direction, which is of great significance for efficient transmission in MCN.


Although progressively adjusting from~$\mathcal{L}_1$ to~$\mathcal{L}_5$ can yield geometric structures with more consistent ISL lengths and more uniform distribution, the effects of geometric layout and topological connectivity on network performance are mutually coupled, and changes in lattice parameters also alter the availability of ISLs. Therefore, it is necessary to subsequently construct a joint optimization model under SML that comprehensively balances objectives such as ISL reliability and delay.

\section{MCN Structure Design Based on SML Paradigm}
\label{sec:pro}

ISL reliability and end-to-end propagation latency are two core metrics for evaluating MCN performance. The former reflects the network's ability to maintain stable paths in the face of dynamics in ISLs, while the latter characterizes the transmission efficiency of packets in networking. Accordingly, this section focuses on the Reliable and Low-latency MCN Design (ReLMD) problem.

Based on the aforementioned \sml~paradigm, this chapter transforms the original network structure design problem into a motif--lattice tuple search problem over the motif space and the lattice space. 
Specifically, this section first establishes an ISL reliability model to evaluate the stability of different local connectivity patterns; then analyzes the impact of geometric layout on end-to-end transmission efficiency and constructs a computable low-latency proxy metric; and finally provides a formal definition of the joint optimization problem on this basis. The overall objective is to achieve an effective trade-off between high link reliability and low transmission latency through the joint optimization of topological connectivity and spatial layout under given constraints.

\subsection{ISL Reliability in Dynamics}

The reliability of optical ISL is affected not only by random factors such as illumination and ATP errors, but is also more directly constrained by the relative motion between satellites. The stronger the relative dynamics between the two satellites at the ends of a ISL, the higher the link interruption probability and the longer the recovery time \cite{Wu2014Correction}, causing ISLs to exhibit pronounced intermittency and instability.

\begin{figure}[t!]
\centering
\includegraphics[width=0.95\linewidth]{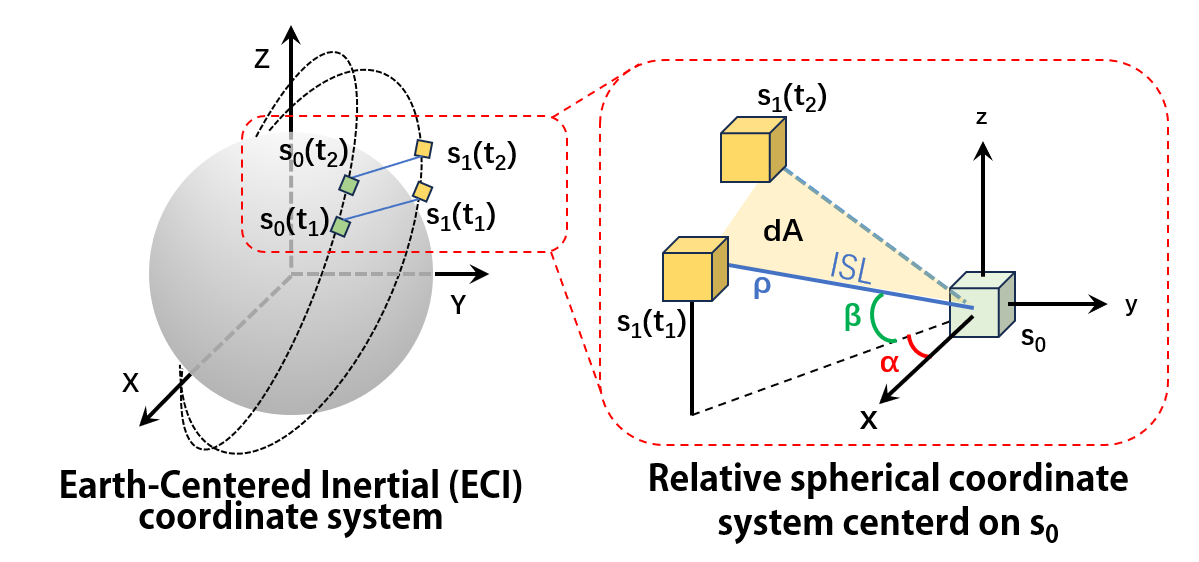}
\caption{The area sweep rate is characterized by the time derivative of the differential area element of the inter-satellite link}
\label{fig:asr}
\end{figure}
To quantify the dynamic characteristics of ISLs, this section introduces the area swept ratio (ASR) $\eta_{e}(t)$ as a metric of the dynamics of link~$e$. As shown in \fig\ref{fig:asr}, the left subfigure depicts the inter-plane ISL established between two satellites (blue line segment), while the right subfigure characterizes the relative motion of the same ISL from the perspective of satellite~$v_0$.
This metric is based on a relative spherical coordinate system centered at one of the two satellites, which is defined in the Earth-centered inertial (ECI) frame. The coordinate components $(\rho, \alpha, \beta)$ denote the radial distance, azimuth, and elevation between the two satellites, respectively. Geometrically, the ASR is defined as the differential spherical area element swept by the line segment connecting the two satellites per unit time, i.e.,
\begin{equation}
\eta_{e}(t) = \frac{dA}{dt} = \frac{\rho^2 \cos (\beta) \cdot d\alpha \cdot d\beta}{dt}
\label{eq:asr}
\end{equation}
where $\alpha$ is the azimuth, $\beta$ is the elevation, and $\rho$ is the relative distance. A smaller ASR indicates weaker relative dynamics of the established ISL and hence generally higher reliability, and vice versa.

On this basis, a stochastic reliability model based on the ASR is further established, such that the larger the ASR of an ISL, the higher its instantaneous interruption probability and the longer its recovery time after interruption. Specifically, the interruption probability of ISL~$e_i$ is modeled as:
\begin{eqnarray}
	P(Z_{e_i} = 0 \mid \eta_{e_i}(t)) = 1 - e^{-\lambda \eta_{e_i}(t)}
\end{eqnarray}
where $Z_{e_i}(t)\in\{0,1\}$ is the binary state of ISL~$e_i$ at time~$t$ (1 for connected and 0 for interrupted), and $\lambda>0$ is a sensitivity parameter determined by the equipment/pointing control capability, which characterizes the influence strength of the ASR on the interruption probability.
After an ISL is interrupted, the laser terminal activates its ATP system to restore the optical ISL. Let the recovery time $y_{e_i}(t)$ be positively correlated with the ASR, i.e.,:
\begin{eqnarray}
	y_{e_i}(t) = Y_{\min} + (Y_{\max}-Y_{\min}) \cdot P(Z_{e_i} = 0 \mid \eta_{e_i}(t))
\end{eqnarray}
where $Y_{\min}$ and $Y_{\max}$ are the preset minimum and maximum recovery times (in seconds), and $y_{e_i}(t)$ denotes the interruption/recovery time scale required after an interruption occurs at time~$t$.
Since an ISL has only two states, connected and interrupted, which alternate randomly, its evolution can be modeled as a typical alternating renewal process\cite{Gallager2013}, with the state transition equation defined as:
\begin{equation}
    Z_{e_i}(t+\Delta t) = \left\{
    \begin{aligned}
		0,~&\text{if}~  Z_{e_i}(t) = 1 ~\text{and}~  r < P(Z_{e_i} = 0 \mid \eta_{e_i}(t))  \\
		0,~&\text{if}~  Z_{e_i}(t) = 0 ~\text{and}~  t-t^\prime < y_{e_i}(t) \\
        1,~&\text{otherwise} \\
    \end{aligned}
    \right.
\end{equation}
where $\Delta t$ is the time step of the discrete simulation, $t^\prime$ denotes the time at which the current interruption begins (i.e., the switching moment when $Z_{e_i}$ changes from 1 to 0), and $r\sim U[0,1]$ is a uniformly distributed random number.
Finally, the reliability $R_e$ of an ISL is quantified by the average proportion of connected time within the observation time window (i.e., the simulation time) $T$, i.e.,:
\begin{eqnarray}
    R_e(e_i) =  \frac{1}{T}\int_{0}^{T} Z_{e_i}(t)\,dt
    \label{eqt:reliability}
\end{eqnarray}
Thus, the differences in the dynamic characteristics of different ISLs are ultimately reflected in their reliability differences. Therefore, the key to achieving a highly reliable network lies in preferentially establishing ISLs with minimal dynamics.

\subsubsection{Low-Latency Layout Design}
\label{sec:sml:ps:layout}

One of the core objectives of \pro~is to design a mega-constellation network with better transmission efficiency, i.e., to generate a network structure such that, under random traffic conditions, the average path of a packet detours less and has a shorter physical propagation distance, thereby effectively reducing the end-to-end delay.
To this end, the following proposition is presented to establish the connection between the geometric layout and the transmission efficiency:

\noindent\textbf{Proposition 1:}
For a Walker configuration MCN with a fixed topology, if its geometric layout makes the average length of the ISLs shorter, then the shortest paths between any two satellites will, on average, detour less and have shorter propagation distances.

\noindent\textbf{Proof:}
Consider the set of all traffic flows $\mathcal{T}$. For any traffic flow $\tau\in\mathcal{T}$, let $\mathcal{P}_\tau$ denote the routing path between its source satellite and destination satellite, which can be viewed as an ordered set of ISLs; $\mathcal{E}$ denotes the set of ISLs in the network. Let $l(e)$ be the physical length of link $e\in\mathcal{E}$; then the propagation distance of path $\mathcal{P}_\tau$ is:
\begin{eqnarray}
l(\mathcal{P}_\tau) = \sum_{e \in \mathcal{P}_\tau} l(e),
\end{eqnarray}
Let $f(e)$ denote the proportion of all traffic flows that pass through ISL $e$, i.e.,:
\begin{eqnarray}
f(e) = \frac{1}{|\mathcal{T}|} \sum_{\tau \in \mathcal{T}} [e \in \mathcal{P}_\tau]
\end{eqnarray}
where $[\cdot]$ is the Iverson bracket, which equals 1 if the statement inside the brackets is true, and 0 otherwise.
Then, for the traffic set $\mathcal{T}$, the average path length $\bar{l}_{\mathcal{P}}$ over all traffic flows can be written as:
\begin{eqnarray}
         \nonumber
         \bar{l}_{\mathcal{P}} &=& \frac{1}{|\mathcal{T}|} \sum_{\tau \in \mathcal{T}} l(\mathcal{P}_\tau) \\
         &=& \frac{1}{|\mathcal{T}|} \sum_{\tau \in \mathcal{T}} \sum_{e \in \mathcal{E}} l(e) \, [e \in \mathcal{P}_\tau] \\
         \nonumber
         &=& \sum_{e \in \mathcal{E}} l(e) \cdot \left( \frac{1}{|\mathcal{T}|} \sum_{\tau \in \mathcal{T}} [e \in \mathcal{P}_\tau] \right)\\
         \nonumber
         &=& \sum_{e \in \mathcal{E}} l(e) \, f(e).
     \end{eqnarray}
This indicates that the average traffic path length $\bar{l}_{\mathcal{P}}$ can be expressed as a weighted sum of ISL lengths, where the weights are determined by the frequency with which a ISL is traversed by the shortest paths.

In practice, traffic demand is geographically highly non-uniform, being concentrated in regions with high population or GDP \cite{roughan2005simplifying}, and varies over time. Therefore, at any given moment, $f(e)$ is often non-uniform as well. 
However, the periodic motion of MCN together with the Earth's rotation produces a long-term averaging effect: on the one hand, each ISL periodically sweeps across regions with non-uniform traffic demand; on the other hand, after an ISL has carried a high load in a certain region, other ISLs in subsequent orbits will also serve the same region at different times.
Therefore, under the combined action of the above periodic effects, the proportions of traffic carried by the ISLs tend to become equal over long-term operation, i.e., for any $e \in \mathcal{E}$, $E[f(e)] \approx \bar{f}_\mathcal{E}$. Consequently, the expected average path length can be expressed as:

\begin{eqnarray}
\label{eqt:propto}
E[\bar{l}_{\mathcal{P}}] = \sum_{e \in \mathcal{E}} l(e) \cdot E[f(e)] = |\mathcal{E}| \cdot \bar{l}_\mathcal{E} \cdot \bar{f}_\mathcal{E}
\end{eqnarray}
where $\bar{l}_\mathcal{E}$ denotes the average length of the ISLs in the set $\mathcal{E}$, $|\mathcal{E}|$ denotes the number of ISLs, and $\bar{f}_\mathcal{E}$ denotes the average of $E[f(e)]$ over all ISLs. A more detailed discussion of the traffic characteristics is provided in \app\ref{app:traffic}.

It can thus be seen that, for a given fixed topology, $\bar{f}_\mathcal{E}$ and $|\mathcal{E}|$ are both constants; therefore, the average path length $E[\bar{l}_{\mathcal{P}}]$ is proportional to the average ISL length $\bar{l}_\mathcal{E}$. 
This proportionality suggests that the transmission efficiency is governed by the spatial compactness of the layout: when the constellation geometry approximates a Delaunay triangulation, each satellite is connected to its nearest spatial neighbors, thereby avoiding overly long edges and producing a locally uniform, compact mesh, which tends to minimize $\bar{l}_\mathcal{E}$ and, in turn, the average path length\cite{giuliari2020internet,li2002distributed}.
Therefore, $\bar{l}_\mathcal{E}$ can serve as a practical and easily computable surrogate for the average end-to-end path length $\bar{l}_{\mathcal{P}}$, providing a simplified yet effective optimization basis for efficient geometric layout design.

\subsection{Reliable and Low-latency Constellation Network Design Problem}

Combining the ISL reliability model and the low-latency layout analysis presented above, this section further formulates the Reliable and Low-Latency MCN Design, (\pro) problem.
Unlike modeling approaches that directly rely on the MCN topology and constellation parameters, this complex problem is simplified in two respects using the SML paradigm:
First, since the global structure is formed by periodically replicating identical motifs in the network, the reliability of the ISLs within a motif can be used to approximate that of the corresponding ISLs in the entire network, thereby significantly reducing the computational complexity of reliability evaluation;
Second, according to the proportional relationship between the average path length and the average ISL length revealed by \eqt\ref{eqt:propto}, the average end-to-end path length, which is difficult to optimize directly, can be converted into the optimization of the average ISL length $\bar{l}_\mathcal{E}$, which serves as a low-latency proxy.
Based on the above simplifications, the \pro problem can be formulated as:
\begin{eqnarray}
    \nonumber
    \mathop{max}\limits_{\mathcal{M}_q,\mathcal{L}_p}~ w_1\cdot  \underbrace{ \left(\dfrac{1}{|\mathcal{E}^{\mathcal{M}_q}|} \sum_{e \in \mathcal{E}^{\mathcal{M}_q}} R_e(e\mid \mathcal{L}_p)\right) }_{\text{high reliability}}+\\
    w_2 \cdot \underbrace{\left( \dfrac{1}{\bar{l}_\mathcal{E} (\mathcal{M}_q,\mathcal{L}_p)}\right) }_{\text{low latency}}
\end{eqnarray}
where $R_e(e\mid \mathcal{L}_p)$ denotes the reliability of ISL $e$ under lattice $\mathcal{L}_p$, and $\mathcal{E}^{\mathcal{M}_q}$ is the set of ISLs contained in motif $\mathcal{M}_q$; $\bar{l}_\mathcal{E}(\mathcal{M}_q,\mathcal{L}_p)$ denotes the average ISL length of the network under a given motif and lattice, whose reciprocal characterizes the low-latency objective;
Here, $w_1$ and $w_2$ are weighting parameters used to balance the trade-off between reliability and latency. Since the reliability term is bounded in $[0,1]$ (see \eqt\ref{eqt:reliability}) while the low-latency term typically falls within $[5\times 10^{-7},1\times 10^{-6}]$ depending on the ISL lengths in the MCN, the two terms differ greatly in magnitude, so that $w_1$ and $w_2$ are introduced to balance their relative weights (in practice, $w_1=1$ and $w_2=10^6$).
This optimization problem is subject to the following constraints:
\begin{eqnarray}
\forall X(v_{x,y}) \in \mathcal{M}_q, \; |X(v_{x,y})|\leq c_{\max}\\
 \mathcal{L}_p\in  \{\mathcal{L}_1,\mathcal{L}_2,\cdots,\mathcal{L}_5\}=\mathbb{L}
\end{eqnarray}
where $X(v_{x,y})$ denotes the connection pattern of satellite node $v_{x,y}$ at position $(x,y)$ in motif $\mathcal{M}_q$, i.e., its connection relationships with other satellite nodes; $|X(v_{x,y})|$ denotes the number of connection vectors in this connection pattern; $c_{\max}$ is the preset maximum number of connections, reflecting the maximum number of ISLs that each satellite can support; and $\mathbb{L}$ is the predefined set of lattices, which, according to the two-dimensional Bravais lattice theory, contains all admissible periodic geometric layouts.

Thus, the complex network structure design problem is transformed into a combinatorial optimization problem within the design space defined by \sml. Its optimal solution corresponds to an optimal motif--lattice combination $(\mathcal{M},\mathcal{L})$ that can achieve a systematic optimal trade-off between ISL reliability and transmission efficiency under the given engineering constraints.

\section{Progressive Lattice and Motif Search Algorithm}%
\label{sec:ps:alg}

The \pro~problem is essentially a combinatorial optimization problem defined over the motif searching space and the lattice searching space, and its global optimum can in theory be obtained by exhaustively enumerating all feasible motif and lattice configurations. However, since a motif is constructed by hierarchically combining connection vectors and connection patterns, the search space expands rapidly as the structural complexity increases, making exhaustive search computationally prohibitive.

To this end, a progressive motif-lattice search algorithm (Progressive Lattice And Motif Search, PLAMS) is proposed. The algorithm explores motif structures of increasing complexity by progressively expanding the connection vector set $\Phi$, and completes the joint search in combination with lattice traversal. Its design is based on the following two core strategies:

\begin{itemize}
	\item \textbf{Depth-first progressive search}: the connection vector set is expanded progressively, candidate vectors are introduced in ascending order of vector length, and a depth-first strategy is adopted for the search, so that the generated network structure evolves from simple to complex;
	\item \textbf{Hierarchical pruning}: during the hierarchical combination of connection vectors, connection patterns, and motifs, constraints are applied and pruning is performed as early as possible at each level by exploiting the properties of depth-first traversal.
\end{itemize}

The \solver~algorithm avoids storing all candidate solutions simultaneously through depth-first traversal, and its progressive search property allows the algorithm to terminate at any time and return the current best solution. Meanwhile, the hierarchical pruning strategy ensures that only a limited number of valid candidate solutions enter the costly evaluation stages, such as constellation construction and orbit simulation, thereby significantly reducing the overall computational overhead.
Starting from the initial constellation state, the algorithm iteratively and jointly optimizes the lattice structure and motif configuration, ultimately yielding a feasible \sn~design that maximizes the objective function as much as possible. As shown in \alg\ref{alg:plams}, the \solver~framework consists of three stages: initialization, search, and evaluation.

\subsection{Initialization}

In the initialization stage, a restricted search space is first constructed based on the parameters of the initial constellation, including the number of orbital planes $N_p$, the number of satellites per plane $M_p$, the phase factor $F$, and the orbital inclination $\theta$. This search space simultaneously characterizes the feasible ranges of the lattice structure and the motif configuration.
Among them, the lattice space is restricted to the five two-dimensional Bravais lattices that detailed in \Sec\ref{sec:lattice}, whose underlying parameters are reconfigured from the initial constellation by function \texttt{LatticeSpace()} (line 1). 
The motif space is determined by physical constraints: the connection vector set $\Phi$ (line 2) specifies the allowed connection patterns, and the geometric template set $\mathcal{R}$ (line 3) specifies the relative positions of the connection patterns within a motif. Together, these definitions determine the feasible space for motif construction.

\subsection{Candidate Solution Search}
Candidate solutions are generated by jointly traversing the lattice space and the motif space. Since only five basic layouts ($\mathcal{L}_1\sim \mathcal{L}_5$) remain in the lattice after parameter reconfiguration, the lattice space can be enumerated directly; in contrast, the motif space is much larger and must be explored using a hierarchical progressive search strategy, which specifically includes the following three levels:

\begin{algorithm}[t!]
    \caption{Progressive Lattice And Motif Search (PLAMS)}
    \label{alg:plams}

    \LinesNumbered 
    \KwIn{initial parameters $N_p,M_p,F,\theta$;~maximum motif size $d_{max}$;~
		maximum connection vector count per pattern $c_{max}$;  }
    \KwOut{optimal structure parameters $N_p^\star,M_p^\star,F^\star,A^\star $}
    $ \mathbb{L} \gets \texttt{LatticeSpace}(N_p,M_p,F,\theta)$;\\ 
	$\Phi \gets \{\phi(0,1),\phi(1,0),\phi(1,-1),{}$\\
	\phantom{$\Phi \gets \{$}$\phi(1,1),\phi(1,2),\phi(1,-2) \}$;\\
	 $\mathcal{R} \gets  \{v_{x,y},v_{x+1,y},v_{x,y+1},v_{x+1,y+1}\}$;\\
	 $Obj^\star \gets 0$;\\
	\For{$\mathcal{L}_p \in \mathbb{L}$ }{
		\For{connection vectors~$\Phi_i \in$ \texttt{Subsets($\Phi$)}}{
			\For{connection patterns~$\mathcal{X}_j \in \bigcup\limits_{c=1}^{c_{max}} \binom{\Phi_i}{c}  $}{
				\textbf{if}   \texttt{IsVisited($\mathcal{X}_j $)} or $|\mathcal{X}_j|==1$ \textbf{then} \textbf{continue};\\
				\For{ $ \mathbf{X}_k,\mathbf{R}_k \in \bigcup  \limits_{d=1}^{d_{max}}(\mathcal{X}_j^{d},\mathcal{R}^d)$}{
			
					\textbf{if}   \texttt{IsVisited($\mathbf{X}_k $)} or \texttt{IsVisited($\mathbf{X}^\prime_k$)} \textbf{then} \textbf{continue};\\		$\mathcal{M}_q \gets \mathcal{M}( \mathbf{X}_k ,\mathbf{R}_k )$; \\ 
					\textbf{if~not}  \texttt{IsConnected($\mathcal{M}_q$)} \textbf{then} ~~\textbf{continue};\\
					$\mathcal{S}^\star \gets \mathcal{S}(\mathcal{M}_q,\mathcal{L}_p) $;\\
					\textbf{if}  \texttt{IsTopoDup($\mathcal{S}(\mathcal{M}_q,\mathcal{L}_p)$)} \textbf{then} \\
					~~\textbf{continue};\\
					\If{ $Obj( \mathcal{S}^\star )>Obj^\star $}{
						$Obj^\star \gets Obj(\mathcal{S}^\star )$;\\
						$N_p^\star,M_p^\star,F^\star,A^\star \gets \texttt{parse} (\mathcal{M}_q,\mathcal{L}_p) $;~ \\
						
					}
					
			}
			}

		}
	}
    \Return{$N_p^\star,M_p^\star,F^\star,A^\star $}
    \end{algorithm}

\subsubsection{Connection vector level} (line 7): the subsets $\Phi_i$ of the connection vector set $\Phi$ are expanded in ascending order of vector length by progressive subset generation. The vector length reflects the connection span: the larger the length, the longer the corresponding ISL length generally is. 
The function \texttt{Subsets()} in line 6 is responsible for progressively constructing the subsets of connection vectors $\Phi$ while avoiding duplicate generation.

\subsubsection{Connection pattern level} (line 8): given the current subset $\Phi_i$ and the maximum number of connections $c_{\max}$, \solver~ generates all valid connection pattern sets $\mathcal{X}_j$ (line 8). 
To avoid redundant exploration, previously visited connection patterns are filtered out by the \texttt{IsVisited()} function. 
In addition, connection patterns containing only a single connection vector are pruned, because such patterns are usually insufficient to guarantee network connectivity (line 9).

\subsubsection{Motif level} (line 12): 
for all connection pattern sets $\mathcal{X}_j$, ordered tuples $\mathbf{X}_k$ of length $d$ (i.e., connection pattern sequences, where $d$ denotes the motif size) are generated, together with the matching set templates $\mathbf{R}_k$.
To eliminate redundancy, tuples $\mathbf{X}_k^\prime$ that are equivalent under order permutation or cyclic shift must also be checked; for example, $\mathbf{X}=(X_1,X_2)$ and $\mathbf{X}^\prime=(X_2,X_1)$ represent the same motif, and only one of them is retained (line 11).
Subsequently, motif candidates are constructed by combining the connection patterns $\mathbf{X}_k$ and the geometric templates $\mathbf{R}_k$ (line 12).

To further suppress the combinatorial explosion, the following pruning constraints are applied:

\begin{itemize}

\item \textbf{Single-traversal check}:
the \texttt{IsVisited()} function (lines 9 and 11) is used to identify and prune duplicate connection patterns and motifs as early as possible, marking them as visited if no duplicate exists.

\item \textbf{Intra-motif connectivity check}:
since randomly generated patterns may lead to disconnection within a motif, the \texttt{IsConnected()} function (line 13) is used to verify whether all satellite nodes in the motif are connected through ISLs.

\item \textbf{Topology deduplication}:
different connection pattern tuples may produce the same topological structure. The \texttt{IsTopoDup()} function (line 15) is used to remove such duplicates to ensure the uniqueness of the topology.
\end{itemize}

\subsection{Solution Evaluation and Optimal-Solution Update}

In each iteration, the objective value $Obj(\mathcal{S}^\star)$ of the current candidate MCN structure $\mathcal{S}^\star$ is evaluated through constellation simulation (line 18). This objective function jointly considers the average ISL reliability and the average ISL length, thereby striking a trade-off between connection robustness and communication delay.
If the objective value of the candidate solution is better than the current best known solution, the corresponding motif $\mathcal{M}_q$ and lattice $\mathcal{L}_p$ are recorded as the new best solution, and are further resolved into the optimized constellation parameters $(N_p^\star,M_p^\star,F^\star)$ and the ISL adjacency matrix $A^\star$ (lines 17--19); otherwise, the search proceeds to the next candidate solution.

\subsection{Complexity Analysis}
The overall time complexity of \solver~ mainly depends on the size of the motif search space and the evaluation cost of each candidate structure. Let $|\Phi|$ be the total number of connection vectors. When the number of connection vectors selected by each satellite does not exceed $c_{\max}$ (i.e., the number of maintained ISLs does not exceed $2c_{\max}$), the total number of connection patterns can be upper-bounded as
\begin{eqnarray}
    |\mathcal{X}| \leq \sum_{c=1}^{c_{\max}} \binom{|\Phi|}{c} = O(|\Phi|^{c_{\max}})
\end{eqnarray}
If the motif size does not exceed $d_{\max}$, the total number of possible motifs can be further upper-bounded as
\begin{eqnarray}
    |\mathcal{M}| = O(|\mathcal{X}|^{d_{\max}}) = O(|\Phi|^{c_{\max}d_{\max}})
\end{eqnarray}
Therefore, the progressive motif search still has exponential complexity in the worst case. However, its hierarchical search structure enables multi-stage early pruning by combining topological and physical constraints, discarding invalid or clearly inferior branches before the candidate structures are fully instantiated, thereby concentrating the computational resources on more promising search paths.

\section{Simulation and Evaluation}
\label{sec:exp}

In this section, PLAMS is employed to search for optimal motif-lattice configurations for four public MCNs (\Sec\ref{sec:exp}-\ref{sec:exp:opt}). Subsequently, the optimization outcomes are evaluated in terms of \udm{MCN reliability} and traffic latency (\Sec\ref{sec:exp}-\ref{sec:exp:eva}). 
We use SNK \cite{snk}, an open source tool to conduct our simulation and evaluation.

\subsection{Simulation Setup}

\noindent\textbf{Constellations.}
We apply Kuiper~\cite{kuiper}, OneWeb~\cite{oneweb}, Telesat~\cite{telesat}, and Starlink~\cite{starlink} as benchmark constellations for optimization. All scenarios are built on their FCC files.
Since constellation operators do not disclose their actual topologies, the ``+Grid'' and ``*Grid''\footnote {Each satellite establishes ISLs with its 4 or 6 nearest neighbors, forming a regular network topology.} connection patterns are adopted to above constellations as baseline topology configurations.
We also include a structure from prior works \cite{basak2025leocraft} that maximizes phase-offset and orbit-count (MPO) to achieve maximal path diversity, adopting it as an additional baseline for comparison.

\noindent\textbf{Traffic model.}
In contrast to non-uniform traffic models, such as those based on GDP or population density\cite{bhattacherjee2019network,giuliari2020internet}, 
We use 5000 uniformly distributed satellite-to-satellite traffic flows for simulation, as it enables the unbiased evaluation of the intrinsic structure performance of the network. We set the requested bandwidth of each traffic demand to be uniformly distributed between 1 and 5 Gbps.
We simulate the network dynamics over 200 discrete time steps, with each step representing a 100-second interval. 
All the evaluations are adopt the shortest-distance-path routing based on Dijkstra algorithm \cite{cormen2022introduction}.

\begin{table}[t!]
	\caption{Original and reconfigured parameters for four MCNs.}
	\centering
	\renewcommand{\arraystretch}{1.1}
	\setlength{\tabcolsep}{5pt} 
	\scalebox{0.9}{
	\begin{tabular}{cc|cccccc|cccc}
	\toprule[1pt]
	\multicolumn{2}{c|}{MCNs}  & $N_P$ & $M_P$ & $F$ & $i$   & \multicolumn{2}{|c|}{MCNs}   & $N_P$  & $M_P$ & $F$ & $i$      \\ \hline
	\multicolumn{1}{c|}{} & $\mathcal{L}_1$ & 22    & 72  & 0   & 53   & \multicolumn{1}{|c|}{}& $\mathcal{L}_1$ & 12   & 49   & 0    & 87.9   \\
	\multicolumn{1}{c|}{} & $\mathcal{L}_2$ & 22    & 72  & -17 & 53   & \multicolumn{1}{|c|}{}& $\mathcal{L}_2$ & 12  & 49  & -1 & 87.9  \\
	\multicolumn{1}{c|}{\multirow{2}{*}{\rotatebox{90}{~~Starlink}}} & $\mathcal{L}_3$ & 35    & 45  & 8   & 53 & \multicolumn{1}{|c|}{\multirow{2}{*}{\rotatebox{90}{~~OneWeb}}}  & $\mathcal{L}_3$ & 17  & 34  & -1 & 87.9   \\
	\multicolumn{1}{c|}{}      & $\mathcal{L}_4$ & 22    & 72  & -20 & 53                         & \multicolumn{1}{|c|}{}                                           & $\mathcal{L}_4$ & 12  & 49  & 8  & 87.9   \\
	\multicolumn{1}{c|}{}        & $\mathcal{L}_5$ & 38    & 41  & -6  & 53                         & \multicolumn{1}{|c|}{}                                           & $\mathcal{L}_5$ & 18  & 32  & -9 & 87.9   \\
	\multicolumn{1}{c|}{}        & MPO & 72    & 22  & 36  & 53                         & \multicolumn{1}{|c|}{}                                           & MPO & 49  & 12  & 25 & 87.9   \\
	\hline
	\multicolumn{1}{c|}{}                                            & $\mathcal{L}_1$ & 17    & 34  & 0   & 51.9                       & \multicolumn{1}{|c|}{}                                           & $\mathcal{L}_1$ & 40  & 33  & 0  & 50.8   \\
	\multicolumn{1}{c|}{}                                            & $\mathcal{L}_2$ & 17    & 34  & -4  & 51.9                       & \multicolumn{1}{|c|}{}                                           & $\mathcal{L}_2$ & 40  & 33  & 19 & 50.8   \\
	\multicolumn{1}{c|}{\multirow{2}{*}{\rotatebox{90}{~~Kuiper}}}   & $\mathcal{L}_3$ & 21    & 27  & 14  & 51.9                       & \multicolumn{1}{|c|}{\multirow{2}{*}{\rotatebox{90}{~~Telesat}}} & $\mathcal{L}_3$ & 32  & 41  & 6  & 50.8   \\
	\multicolumn{1}{c|}{}                                            & $\mathcal{L}_4$ & 17    & 34  & -12 & 51.9                       & \multicolumn{1}{|c|}{}                                           & $\mathcal{L}_4$ & 40  & 33  & 1  & 50.8   \\
	\multicolumn{1}{c|}{}                                            & $\mathcal{L}_5$ & 23    & 25  & -4  & 51.9                       & \multicolumn{1}{|c|}{}                                           & $\mathcal{L}_5$ & 34  & 38  & -7 & 50.8   \\
	\multicolumn{1}{c|}{}   & MPO & 34    & 17  & 18  & 51.9        & \multicolumn{1}{|c|}{}       & MPO & 88  & 15  & 44 & 50.8   \\
	\toprule[1pt]
\end{tabular}
	}
	\label{tab:public}

\end{table}

\subsection{Optimal Structure Search}
\label{sec:exp:opt}

\begin{figure*}[t!]
    \centering
    \includegraphics[width=1\linewidth]{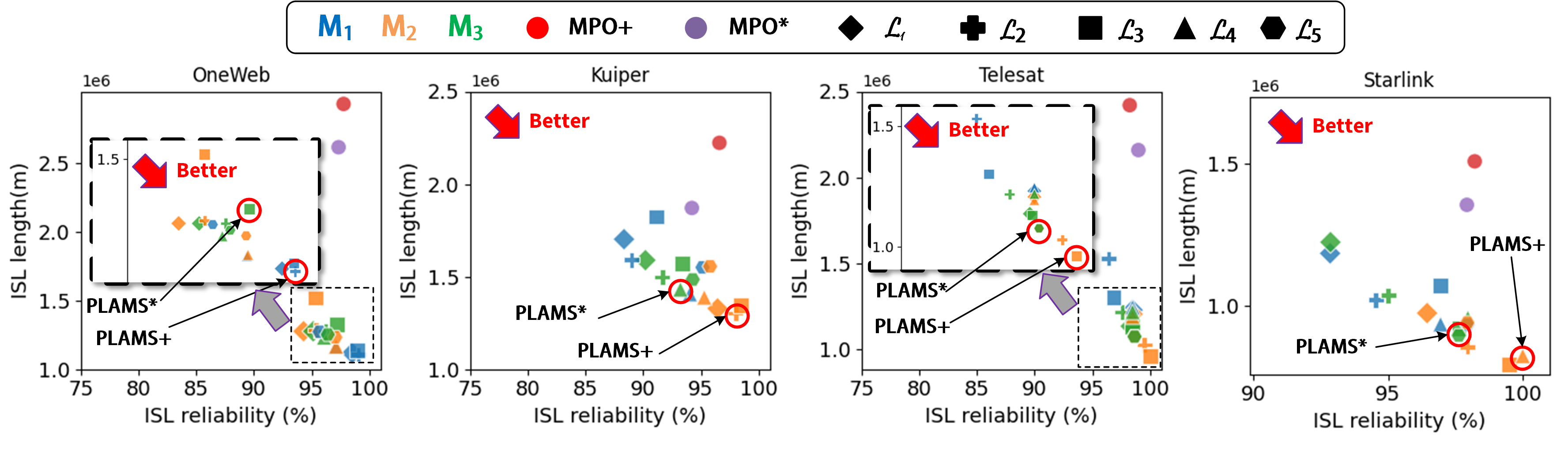}
    \caption{
            Comparison of the average ISL length and ISL reliability of MCN structures under different design methods. Since the $\textit{+Grid}$ and $\textit{*Grid}$ variants are numerous, they are distinguished by different colors and scatter shapes.
    }
    \label{fig:search} 
\end{figure*}

\begin{table*}[t!]
    \centering
    \caption{Optimal motif-lattice combinations of MCN structures under different design methods.}
    \centering
    \renewcommand{\arraystretch}{1.2}
    \scalebox{1}{
        \begin{tabular}{|l|cccc|ccc|}
            \hline
            \multirow{2}{*}{} & \multicolumn{4}{c}{4-ISL}                                           & \multicolumn{3}{|c|}{6-ISL}                                                                                                                                                                                                                                                                                        \\ \cline{2-8}
                              & \multicolumn{1}{c|}{+Grid1}                                         & \multicolumn{1}{c|}{+Grid2}                                         & \multicolumn{1}{c|}{$\textit{MPO}_+$}   & \multicolumn{1}{c|}{$\textit{PLAMS}_+$ }             & \multicolumn{1}{c|}{*Grid}                                          & \multicolumn{1}{c|}{$\textit{MPO}_*$}   & $\textit{PLAMS}_*$              \\ \hline
            OneWeb            & \multicolumn{1}{l|}{\multirow{4}{*}{$\mathcal{L}_1+\mathcal{M}_1$}} & \multicolumn{1}{l|}{\multirow{4}{*}{$\mathcal{L}_1+\mathcal{M}_2$}} & \multicolumn{1}{l|}{\multirow{4}{*}{-}} & \multicolumn{1}{l|}{$\mathcal{L}_2+\mathcal{M}_1$} & \multicolumn{1}{l|}{\multirow{4}{*}{$\mathcal{L}_1+\mathcal{M}_3$}} & \multicolumn{1}{l|}{\multirow{4}{*}{-}} & $\mathcal{L}_4+\mathcal{M}_3$ \\
            Kuiper            & \multicolumn{1}{l|}{}                                               & \multicolumn{1}{l|}{}                                               & \multicolumn{1}{l|}{}                   & \multicolumn{1}{l|}{$\mathcal{L}_2+\mathcal{M}_2$} & \multicolumn{1}{l|}{}                                               & \multicolumn{1}{l|}{}                   & $\mathcal{L}_3+\mathcal{M}_3$ \\
            Telesat           & \multicolumn{1}{l|}{}                                               & \multicolumn{1}{l|}{}                                               & \multicolumn{1}{l|}{}                   & \multicolumn{1}{l|}{$\mathcal{L}_3+\mathcal{M}_2$} & \multicolumn{1}{l|}{}                                               & \multicolumn{1}{l|}{}                   & $\mathcal{L}_5+\mathcal{M}_3$ \\
            Starlink          & \multicolumn{1}{l|}{}                                               & \multicolumn{1}{l|}{}                                               & \multicolumn{1}{l|}{}                   & \multicolumn{1}{l|}{$\mathcal{L}_3+\mathcal{M}_2$} & \multicolumn{1}{l|}{}                                               & \multicolumn{1}{l|}{}                   & $\mathcal{L}_5+\mathcal{M}_3$ \\ \hline
        \end{tabular}}
    \label{tab:search}
\end{table*}

\tab\ref{tab:public} presents the geometric constellation parameters re-adjusted according to the layout in \Sec\ref{sec:lattice-adj}, together with their corresponding lattices~$\mathcal{L}_2\sim \mathcal{L}_5$.
The original layout is~$\mathcal{L}_1$, which serves as one of the baseline structures involved in the comparison.
To reduce computational complexity, the set of connection vectors is restricted to a small set: $\Phi = \{\phi_{(0,1)}, \phi_{(1,0)}, \phi_{(1,-1)}\}$. In addition, the maximum number of connection vectors per satellite is limited to~$c_{\max} = 3$, and the maximum motif size is set to~$d_{\max} = 1$, meaning that only single-satellite motifs are considered.
Under the above restricted search space, PLAMS only needs to search over the combinations of 3 canonical motifs and 5 types of lattices. The candidate motifs are as follows:
\begin{eqnarray}
\nonumber
\mathcal{M}_1(\mathbf{R}_1,\mathbf{X}_1),& \mathbf{R}_1 = (v_{x,y}),&
\mathbf{X}_1= (\{\phi_{(0,1)},\phi_{(1,0)}\})\\
\nonumber
\mathcal{M}_2(\mathbf{R}_2,\mathbf{X}_2),& \mathbf{R}_2 = (v_{x,y}),&
\mathbf{X}_2 = (\{\phi_{(0,1)},\phi_{(1,-1)}\})\\
\nonumber
\mathcal{M}_3(\mathbf{R}_3,\mathbf{X}_3),& \mathbf{R}_3 = (v_{x,y}),&
 \mathbf{X}_3= (\{\phi_{(0,1)},\phi_{(1,0)},\phi_{(1,-1)}\})
\end{eqnarray}
Here, motifs $\mathcal{M}_1$ and~$\mathcal{M}_2$ each contain two connection vectors, corresponding to the satellite with four ISLs connection pattern (4-ISL), whereas motif~$\mathcal{M}_3$ contains three connection vectors, corresponding to the satellite with six ISLs connection pattern (6-ISL).

\fig\ref{fig:search} visualizes all candidate \sn structures as a scatter plot, where each point represents a distinct motif-lattice combination. The vertical axis corresponds to the average inter-satellite link (ISL) length (the shorter, the better), and the horizontal axis represents the ISL reliability (the higher, the better). Since operators have not disclosed the exact topologies of their deployed networks, the baseline structures (+Grid1, +Grid2, and *Grid) are obtained by applying motifs~\(\mathcal{M}_1, \mathcal{M}_2, \mathcal{M}_3\) to the original constellation layout.
It can be observed that the \solver-optimized structures (denoted as \solver+ for the 4-ISL pattern and \solver* for the 6-ISL pattern) achieve a shorter average ISL length and higher reliability than the original baselines (marked as diamonds). In contrast, the MPO-optimized structures (marked as circles) achieve higher ISL reliability, but at the cost of a significantly increased link length. This is because the MPO method tends to increase the number of orbital planes\cite{basak2025leocraft}, thereby leading to longer intra-plane links and shorter inter-plane links.

Although inter-plane ISLs are more dynamic due to the larger range of pointing angle variations, in the \solver-generated structures the shorter ISL lengths reduce the ASR, thereby mitigating this dynamicity and improving ISL reliability.
 \tab\ref{tab:search} summarizes the motif-lattice combinations of the constellation structures obtained by \solver~ and other methods. In the following, the performance of the \solver~method is evaluated against other design methods to demonstrate its effectiveness.

\subsubsection{Network Performance Evaluation}
\label{sec:exp:eva}

Based on the network structures obtained by \pro~during optimization, we comprehensively compare the \solver-optimized, MPO-optimized, and baseline structures, using evaluation metrics including network capacity, throughput, and latency (where the latency metrics comprise the path stretch, hop count, and round-trip time).

\paragraph{Network Capacity and Throughput}

\begin{figure}[t!]
    \centering
    \includegraphics[width=0.9\linewidth]{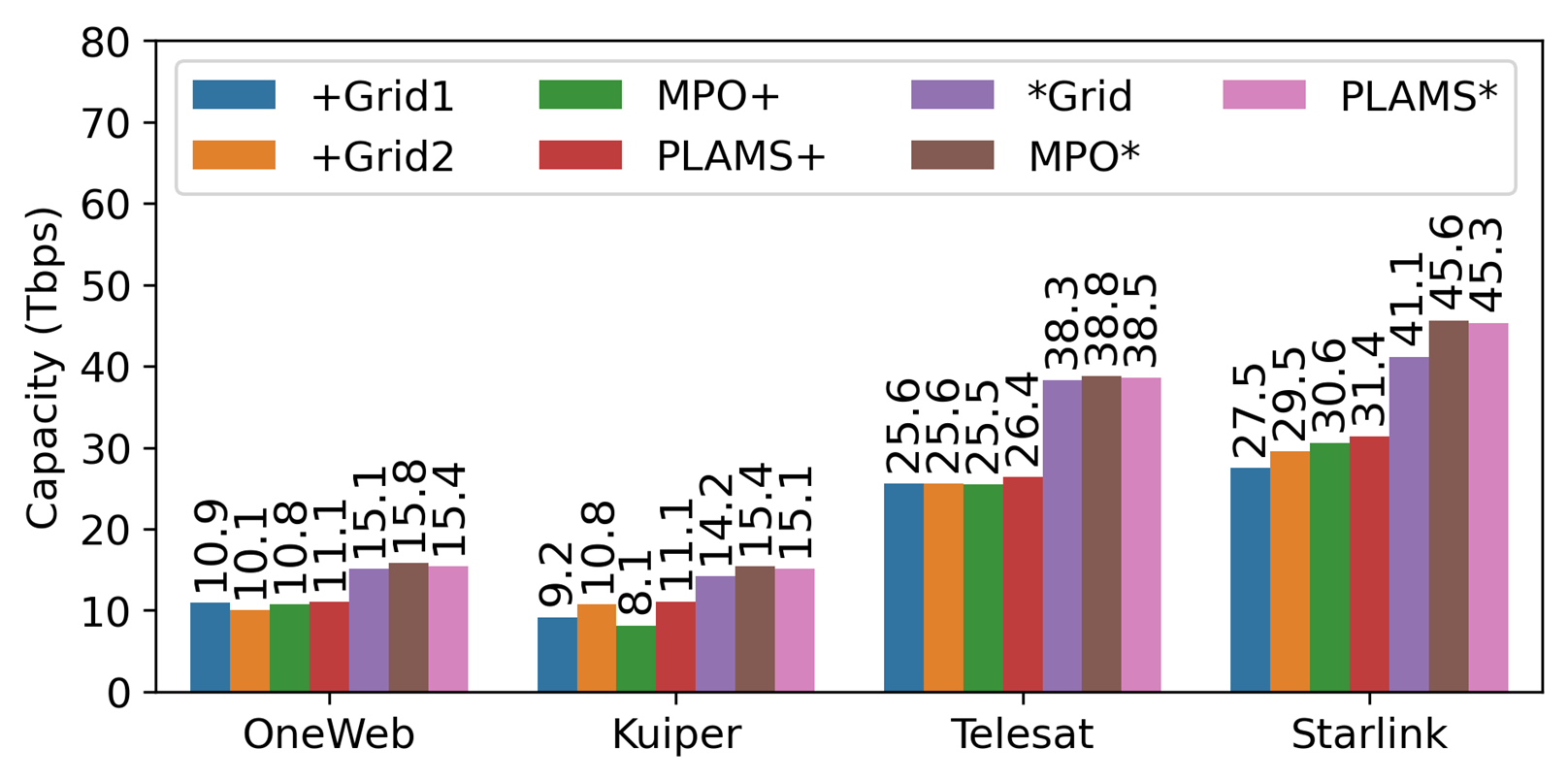}
    \caption{
       Comparison of network capacity under different structures}
    
    \label{fig:exp:cap}
\end{figure}

\fig\ref{fig:exp:cap} shows the network capacity under different network structures. The capacity is estimated using the dynamic network capacity model\cite{wang2026monte}, which assumes that all ISLs are laser links with a fixed capacity of 10~Gbps and sums the instantaneous capacities of all active ISLs.
Among all 4-ISL configurations (+Grid, MPO+, and \solver+), the structures generated by \solver~ achieve the highest capacity (red bars).
By jointly optimizing the network layout and the topological connectivity, \solver~preferentially selects connection patterns with lower dynamics, thereby improving the average ISL reliability and, in turn, the overall network capacity.
In the 6-ISL configurations (*Grid, MPO*, and \solver*), the increased number of ISLs improves the network capacity of all structures. Although \solver outperforms all baseline structures, it is slightly inferior to MPO.
This is because MPO tends to increase the number of orbital planes, thereby leading to longer intra-plane ISLs and shorter inter-plane ISLs. Although inter-plane ISLs are more dynamic due to the larger pointing angle variations, the shorter ISL lengths reduce the ASR. This mitigates the dynamic pressure, improves the average ISL reliability, and ultimately enables MPO to achieve the highest network capacity.
 \begin{figure}[t!]
    \centering
    \includegraphics[width=1\linewidth]{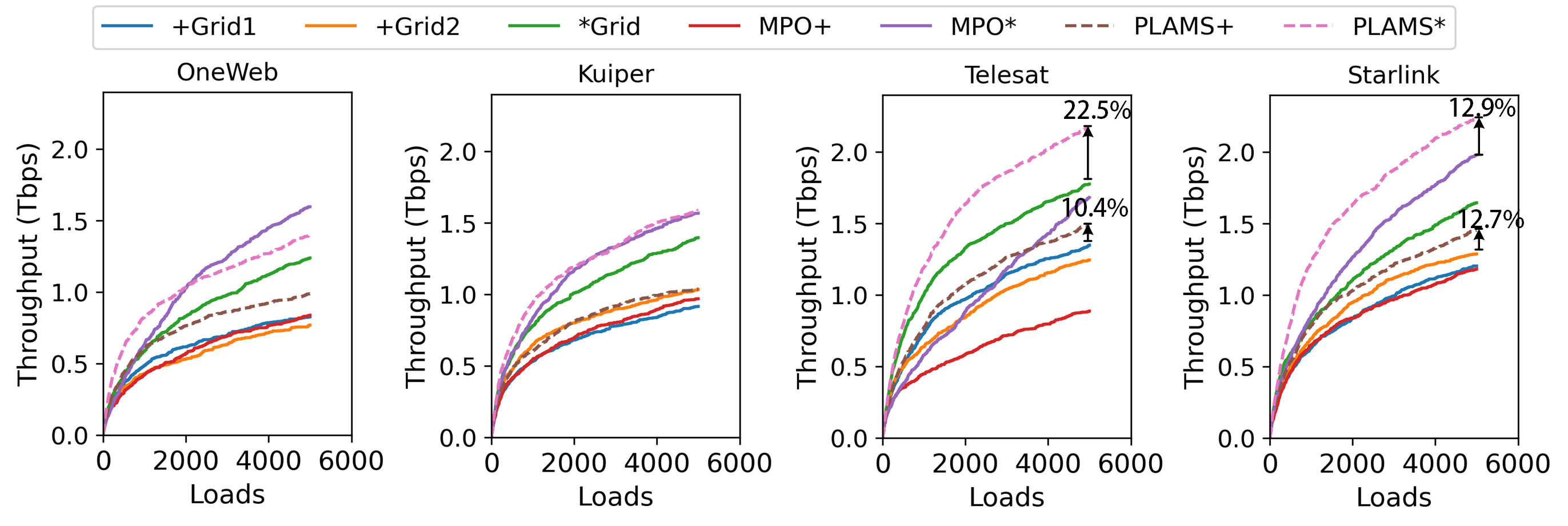}
    \caption{
        Comparison of network throughput under different structures
    }
    \label{fig:exp:thp}
\end{figure}

\fig\ref{fig:exp:thp} evaluates the throughput of the above network structures under the same traffic load.
This section adopts the Monte Carlo method to estimate the throughput\cite{wang2026monte}, which sums the bandwidth allocated to each flow to obtain the aggregate throughput and is applicable to dynamic networks.
It can be observed that in the OneWeb and Kuiper scenarios, the \solver-optimized structures outperform the baseline structures, but are slightly inferior to or comparable with the MPO structures; whereas in the Telesat and Starlink scenarios, the \solver-optimized structures significantly outperform all baseline structures in terms of throughput, improving it by 12.7\% and 22.5\% over the second-best structure in the 4-ISL and 6-ISL cases, respectively.
This performance improvement stems, on the one hand, from the more reliable topological connectivity that increases the network capacity, and on the other hand, from the stronger symmetry of the optimized structures.
The stronger symmetry makes traffic in different directions generally have similar hop counts, makes the distribution of ISLs across the network more balanced, and provides richer choices of link directions.
Consequently, traffic is less likely to concentrate on a few ISLs or satellites, local congestion is alleviated, the network load is distributed more evenly, and the overall throughput is ultimately improved.
 \begin{figure}[t!]
    \centering
    \subfloat[Traffic path stretch]{
    \includegraphics[width=0.95\linewidth]{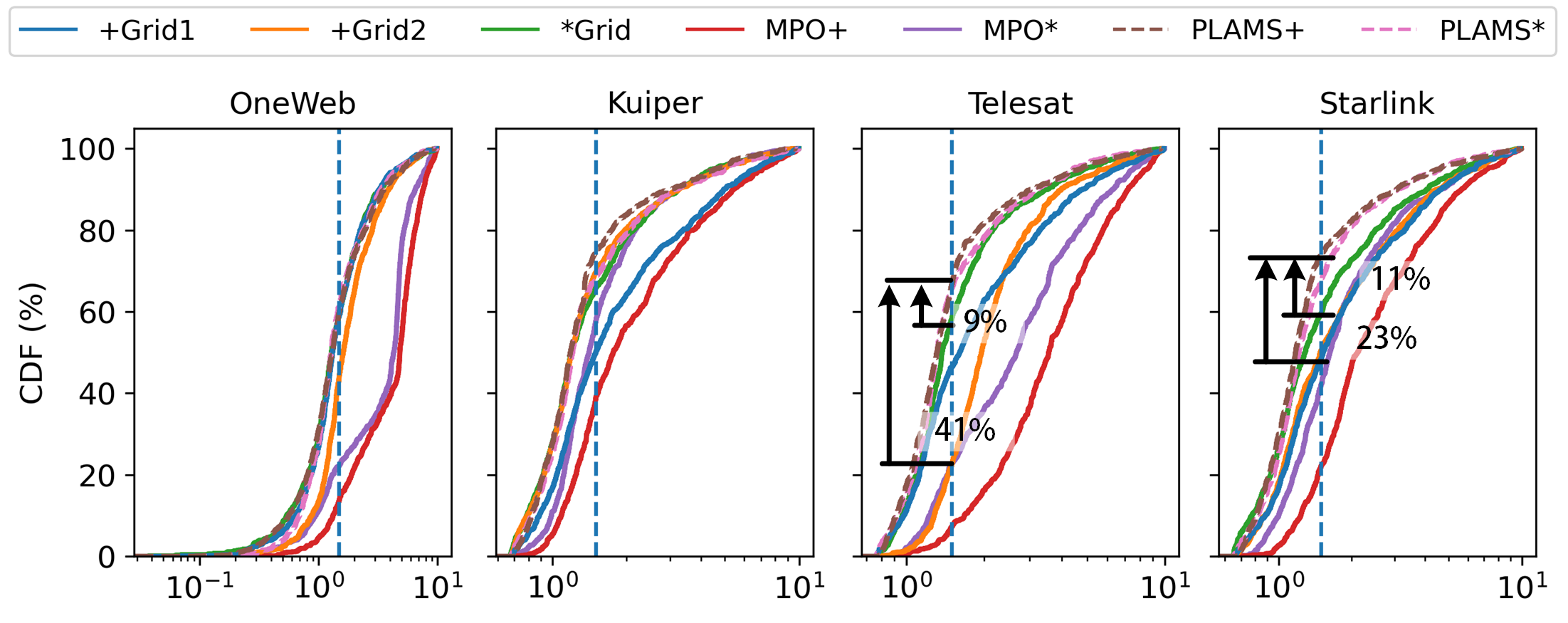}
    }
    \\
    \subfloat[Traffic path hop count]{
    \includegraphics[width=0.95\linewidth]{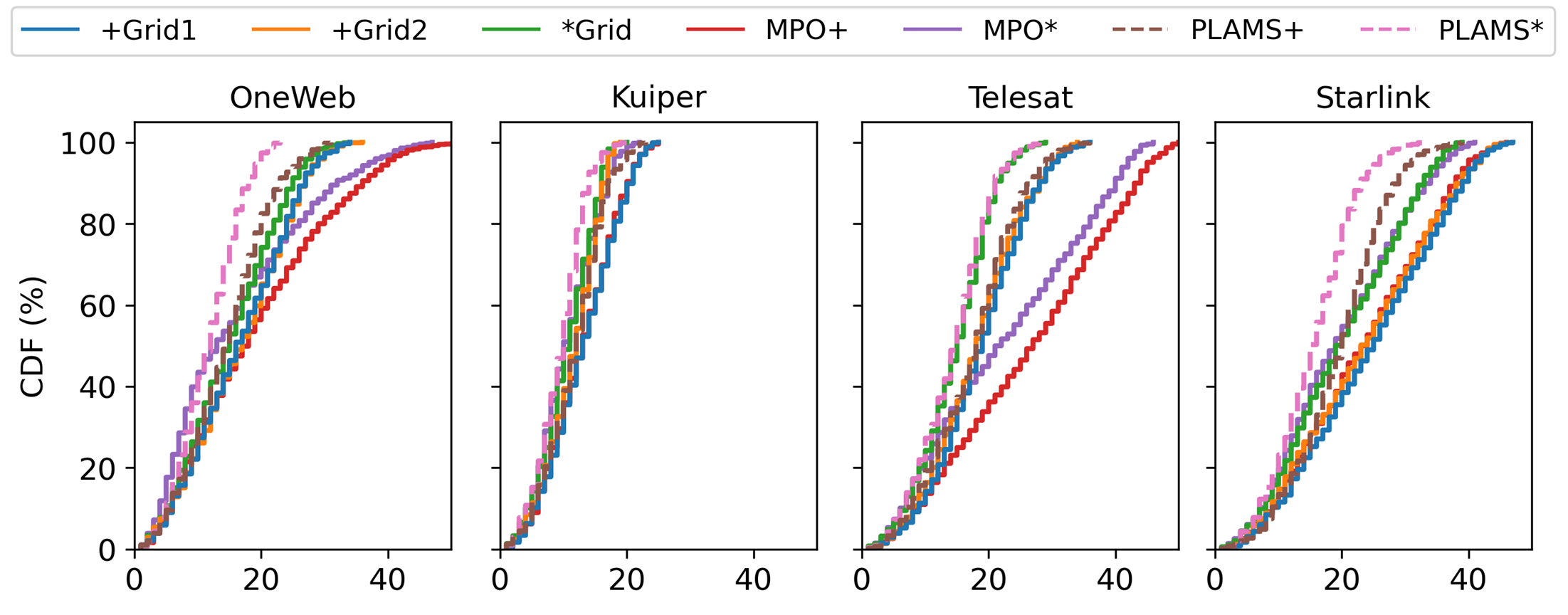}
    }\\
    \subfloat[Round-trip time (RTT)]{
        \includegraphics[width=0.95\linewidth]{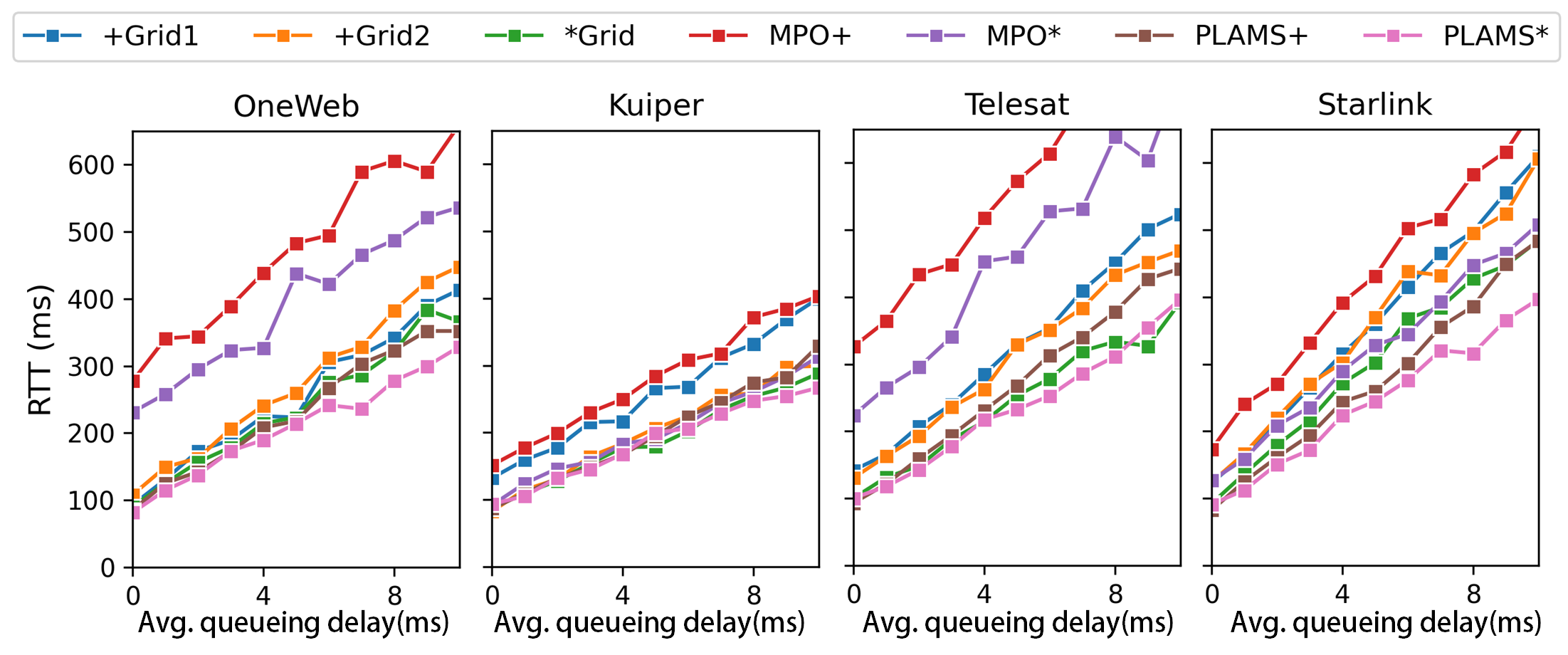}
    }
    \caption{
     Comparison of latency performance between \solver-optimized and baseline structures}
    
    \label{fig:exp:lat}
\end{figure}

\begin{figure*}[t!]
	\begin{center}
		\includegraphics[width=0.8\linewidth]{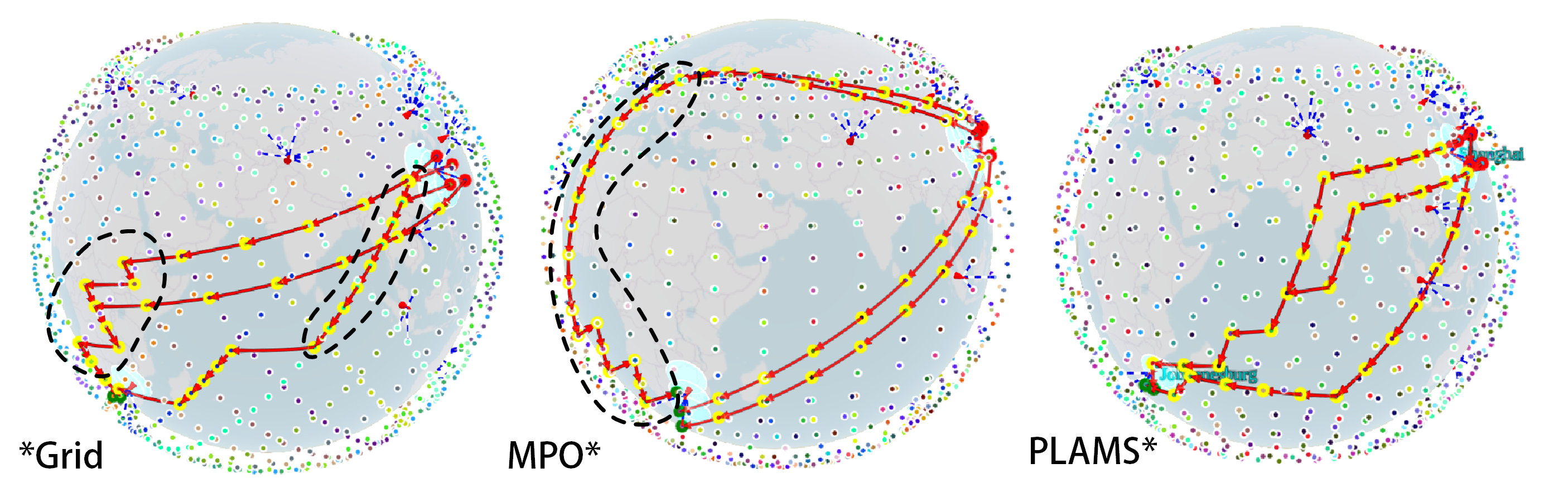}
	\end{center}
	\caption{
        Illustration of end-to-end routing paths from Shanghai to Johannesburg in Starlink under different structures
    }
	\label{fig:exp:pathshow}
\end{figure*}

\paragraph{Latency Performance}

\fig\ref{fig:exp:lat} shows the propagation delay, hop count, and round-trip time (RTT) under synthetic traffic, which together reflect the end-to-end latency performance of the evaluated network structures.
To eliminate the impact of the path-length differences of random traffic on the propagation-delay evaluation, the experiments adopt the ``traffic path stretch'' as a normalized metric, defined as the ratio of the traffic path length to the geodesic distance between the source and destination nodes\cite{bhattacherjee2019network}.
In addition, when evaluating the RTT, it is assumed that end-to-end routing is symmetric and that all satellites have the same packet processing delay (set to 1~millisecond). The RTT is computed by combining the propagation delay and the hop count, and different degrees of network congestion are reflected by the corresponding queueing delay.
\fig\ref{fig:exp:lat}~(a) shows the traffic path stretch, where the blue vertical dashed line indicates a path stretch of $\lambda = 1.5$, corresponding to the lower bound of the end-to-end latency for a direct terrestrial fiber connection.
From the proportion of traffic with a path stretch of $\lambda < 1.5$, it can be seen that in the OneWeb and Kuiper networks, the \solver-optimized structures (dashed lines) perform comparably to or better than the baseline structures.
In the Telesat and Starlink networks, under the 6-ISL configuration the proportion corresponding to the \solver* structures is 9\% and 11\% higher than the baseline, respectively; under the 4-ISL configuration, the proportion corresponding to the \solver+ structures is 41\% and 23\% higher than the baseline, respectively.

\fig\ref{fig:exp:lat}~(b) shows the traffic hop count under different structures. In all constellations, \solver~ achieves the lowest average hop count. Moreover, \solver* under the 6-ISL pattern outperforms \solver+ under the 4-ISL pattern, because its richer distribution of link directions can form paths with less detour.
 \fig\ref{fig:exp:lat}~(c) shows the RTT under different structures. In the Kuiper, Telesat, and Starlink systems, compared with the baseline structures having the same number of ISL configurations, the \solver structures achieve better or comparable RTT performance. In particular, the \solver* structures achieve the lowest RTT in all constellations. In addition, even with a smaller number of ISL configurations, \solver+ outperforms the baseline structures in the OneWeb and Starlink systems.

The above results clearly show that, in various \sn scenarios, the \solver-optimized structures achieve lower latency than the baseline structures. This advantage is attributed to the more symmetric structures produced by \solver: such structures make the ISLs in different directions more balanced, thereby maintaining a low path detour regardless of the traffic direction.
In addition, because the ISL lengths in the \solver-optimized structures remain relatively uniform, the sensitivity of the hop-count distribution to the traffic direction is greatly reduced, ultimately resulting in fewer hops for each flow.

\fig\ref{fig:exp:pathshow} shows the routing paths from Shanghai to Johannesburg in Starlink under the *Grid, MPO-optimized, and proposed \solver-optimized structures. It can be observed that although the *Grid and MPO designs provide some efficient paths, some paths have excessively high hop counts or obvious detours, as indicated by the black dashed lines in the figure. In contrast, the paths in the \solver-optimized structures consistently have fewer hops and shorter detours.

\section{Limitations and Future Work}
\label{sec:limitations}


\noindent\textbf{Exploration of more complex motifs.}
This work considers only a limited set of connection vectors and motifs built from a single connection pattern when optimizing constellations, which yields relatively simple network topologies. This restriction stems from the implementation of \solver, in which every candidate structure must be instantiated as a complete constellation scenario and then evaluated, a process that is computationally time-consuming. 
In future work, we will extend the design space to include network structures composed of more diverse motifs, so as to explore optimal architectures for larger-scale mega-constellations.

\noindent\textbf{Multi-shell MCN design.}
It should be noted that this work focuses exclusively on two-dimensional lattices applied to single-shell constellations. The layout design for multi-layer constellations would require not only three-dimensional lattices but also addressing relative motion induced by differences in orbital altitudes. 
In the future, we will dedicate our research to exploring the integration of diverse motif and lattice combinations for the design of multi-shell constellation networks.


\section{Conclusion}
\label{sec:con}

This study has addressed the issues in designing mega-constellation networks with a new paradigm. To mitigate the high-dimensional complexity in traditional optimization-based approaches, we introduced the SML framework, which models the structure of an MCN as the combination of a lattice and motifs. Within this framework, we formalized the \pro~problem and developed an optimization algorithm to obtain the optimal MCN structure.
Our results demonstrate that optimal MCN performance is achieved when its ISLs are spatially uniform, minimally dynamic, and arranged to approximate a Delaunay triangulation. This finding suggests that a symmetric and uniformly distributed geometric layout is a fundamental design principle for mega-constellation networks, and the SML framework provides a solid foundation for the design of future large-scale MCNs.





\appendix

\subsection{Primitive Transition Vectors}
\label{app:pvec}

To formally establish the mapping from Walker constellation parameters $(N_p,M_p,F,\theta)$ to the five Bravais lattice types, we adopt a two-dimensional planar approximation of the MCN in the vicinity of the equator. Within this approximation, the discrete OPCS is interpreted as a continuous geometric space spanned by two primitive translation vectors, $|\mathbf{a}_1|$ and $|\mathbf{a}_2|$, which jointly describe the periodic arrangement of motifs.

\begin{figure}[htbp]
    \centering
    \includegraphics[width=0.95\linewidth]{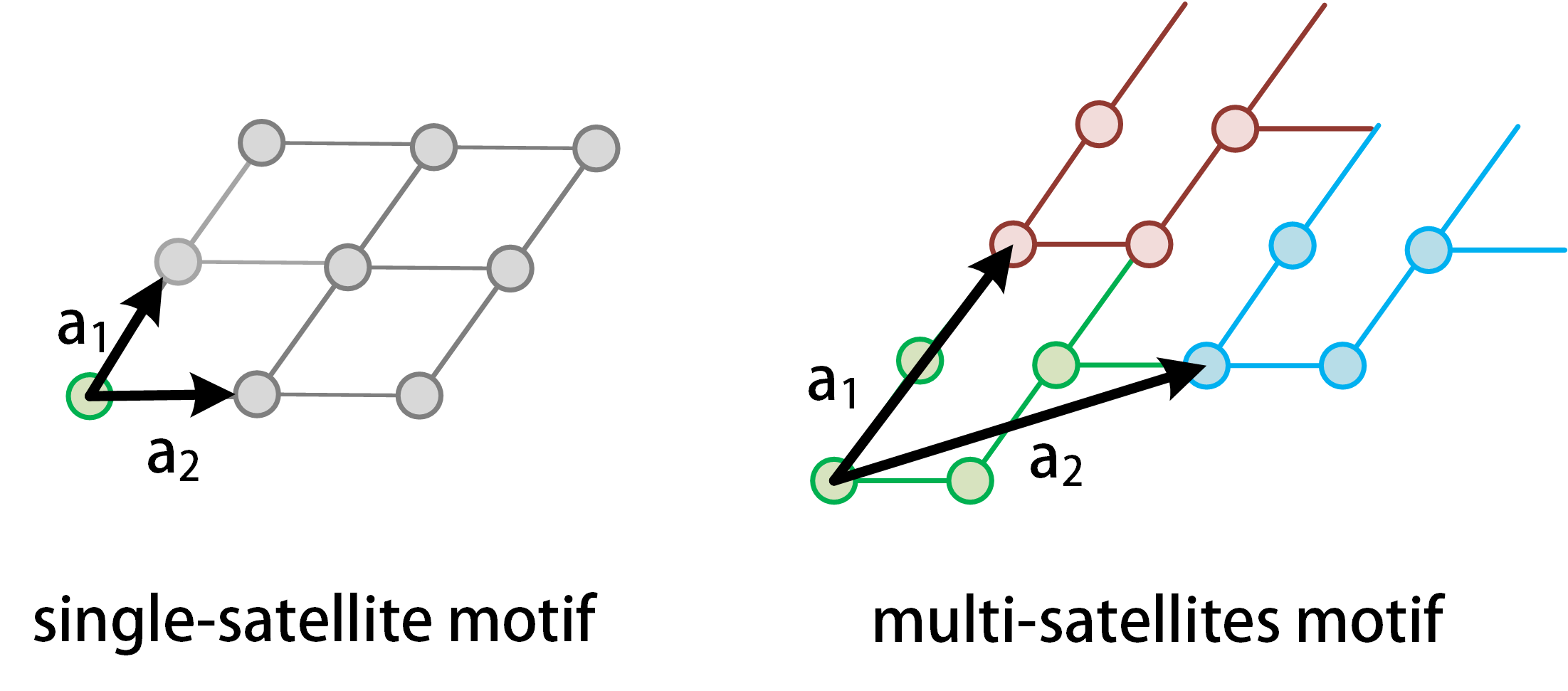}
    \caption{The primitive translation vectors in MCN constructed by single‑satellite motif or multi‑satellite motif.}
    \label{fig:pvec}
\end{figure}

The \fig\ref{fig:pvec} illustrates MCN structures generated by the primitive transition vectors $\textbf{a}_1$ and $\textbf{a}_2$ in the case of single-satellite motif (left panel) and multi-satellites motif (right panel). 
For a single-satellite motif $(d=1)$, $\mathbf{a}_1$ typically aligns with the orbital direction, causing $\mathbf{a}_2$ to align with the inter-orbit ISLs. 
In contrast, for a multi‑satellite motif, (i.e., $d=4$ in right panel), the primitive translation vectors $\mathbf{a}_1$ and $\mathbf{a}_2$ are not constrained to align along the conventional intra‑orbit or inter‑orbit ISL directions. Instead, they are determined according to the geometric definition of a Bravais lattice: the two shortest, non‑parallel vectors that span the periodic repetition of the motif. Consequently, $\mathbf{a}_1$ and $\mathbf{a}_2$ are chosen solely based on the motif’s spatial arrangement, enabling the description of complex periodic connectivity patterns beyond simple orbital alignments.

\subsection{Details in Reconfiguration}
\label{app:reconfig}
In \Sec\ref{sec:lattice-adj}, we proposed a scheme for reconfiguring existing constellations into different lattice layouts and gave the corresponding parameter-adjustment formulas in \tab\ref{tab:lattice}. To clarify the geometric meaning of these formulas, this appendix details the geometric relationships of each lattice layout and presents the derivations of the corresponding parameter-adjustment formulas. In the subsequent figures, red nodes represent the current satellite~$v_{x,y}$; blue and green nodes represent the intra-plane adjacent satellite~$v_{x,y-1}$ and the inter-plane adjacent satellites~$v_{x+1,y}$, $v_{x+1,y+1}$, respectively; and gray nodes represent the position where satellite~$v_{x+1,y}$ would be located under the adjacent plane with phase factor~$F=0$.

\begin{figure}[htbp]
\begin{center}
	\subfloat[$\Delta f/\cos\theta < \Delta \Omega$]{
		\includegraphics[width=0.48\linewidth]{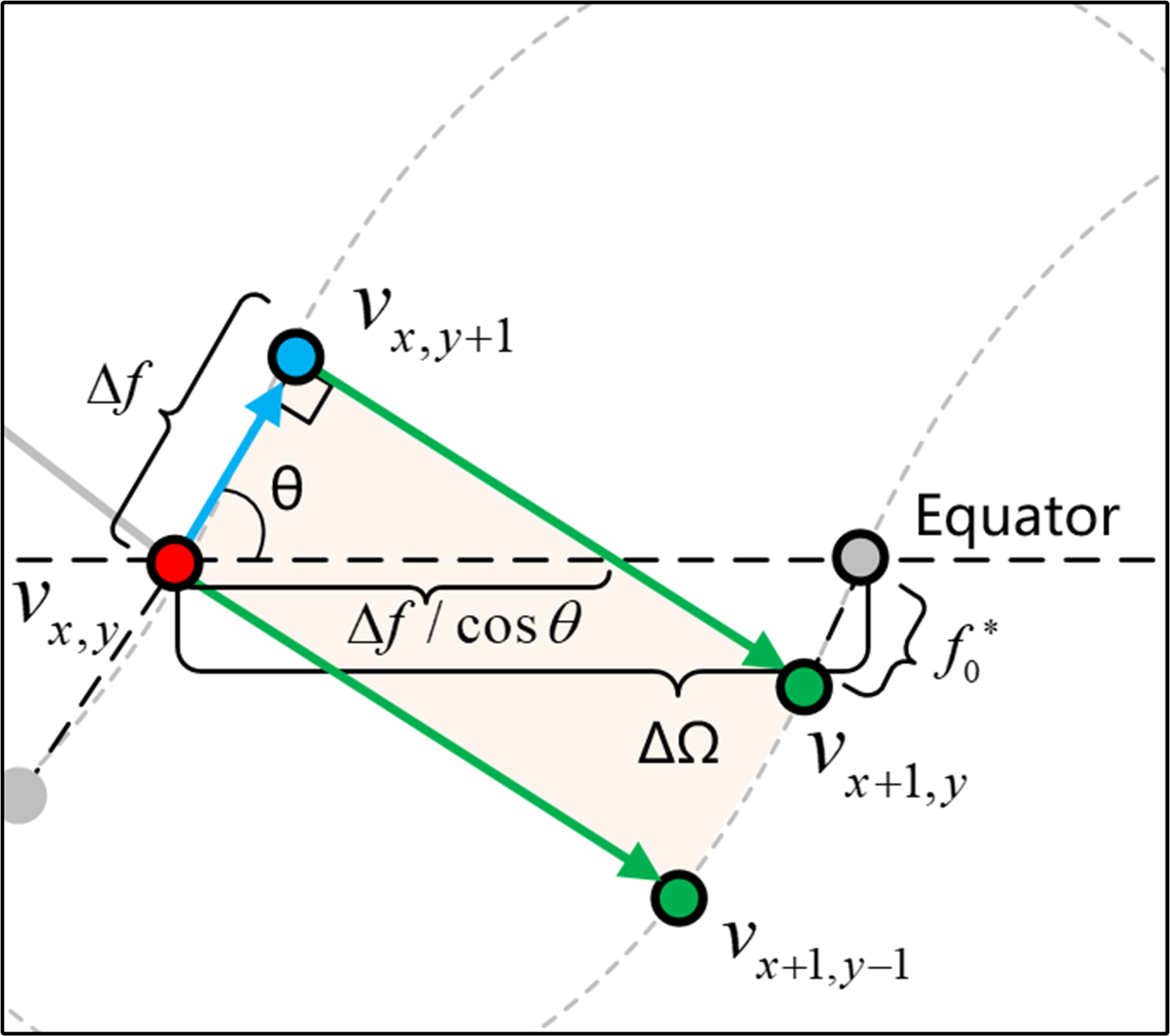}
	}\subfloat[$\Delta f/\cos\theta > \Delta \Omega$]{
		\includegraphics[width=0.48\linewidth]{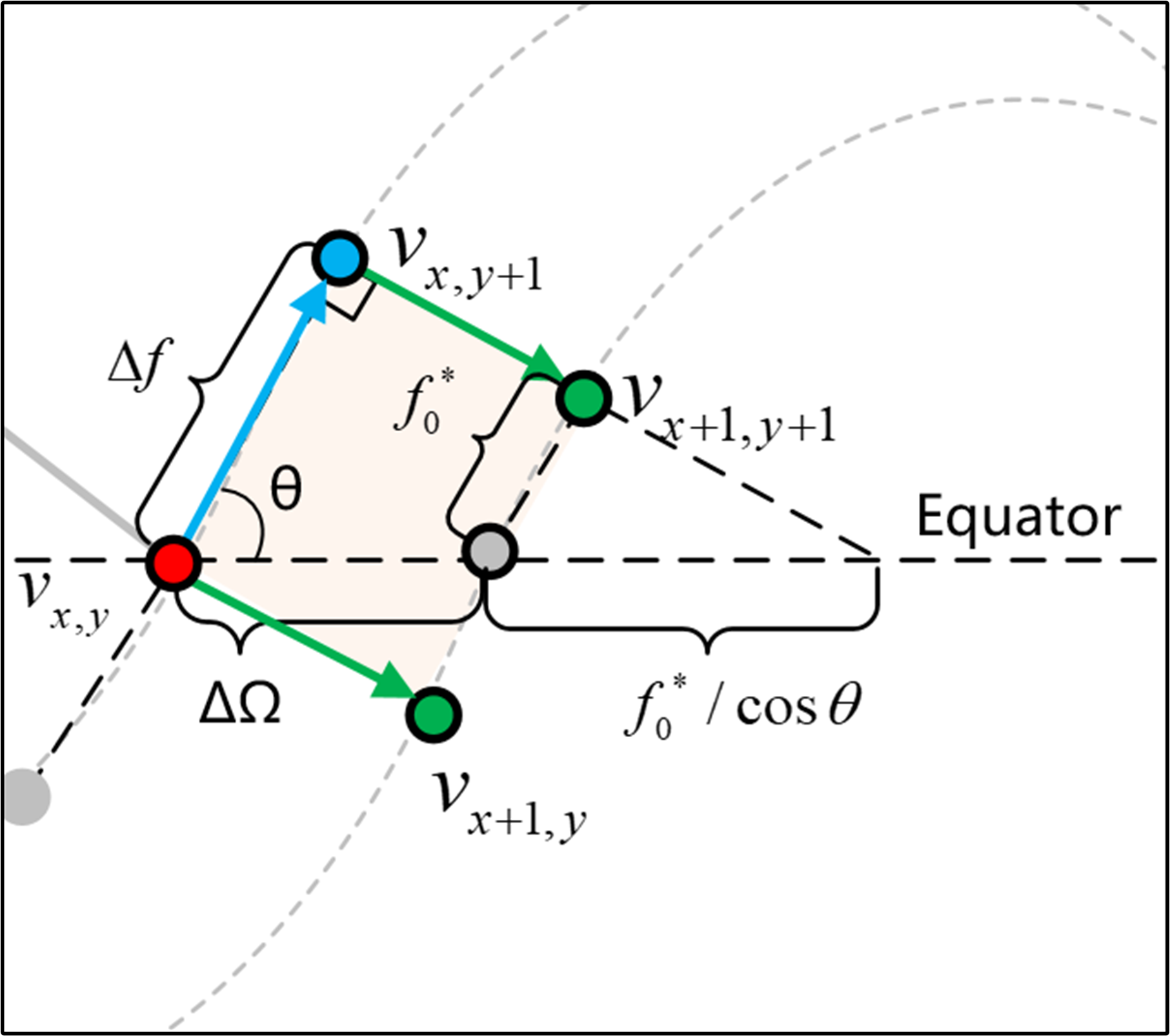}
	}
\caption{Relative geometric configuration of satellites in the~$\mathcal{L}_2$ layout}
\label{fig:l2adj}
\end{center}
\end{figure}

\subsubsection{Reconfiguring to the~$\mathcal{L}_2$ layout}

Reconfiguring to the~$\mathcal{L}_2$ layout only requires changing the phase factor~$F$ of the Walker constellation, so that satellites~$v_{x,y}$, $v_{x,y+1}$, $v_{x+1,y}$, and~$v_{x+1,y+1}$ form an approximately rectangular configuration near the equator. Depending on the relative magnitudes of the RAAN spacing~$\Delta \Omega$ and phase difference~$\Delta f$, two geometric cases exist, as shown in \fig\ref{fig:l2adj}.

When $\Delta f / \cos\theta < \Delta \Omega$, the inter-plane satellite spacing is larger, and the rectangular geometric relationship in \fig\ref{fig:l2adj}~(a) yields:
\begin{eqnarray}
\dfrac{f_0^\star}{\Delta f} = \dfrac{\Delta \Omega - f_0^\star / \cos\theta}{f_0^\star / \cos\theta}
\end{eqnarray}
where $f_0^\star$ is the reconfigured phase difference. From the rectangular geometric relationship, the phase factor can then be obtained as:
\begin{align}
\label{eqt:l2f1}
F^\star &= \dfrac{f_0^\star}{\Delta f} N_p
        = \Bigl( \dfrac{\Delta \Omega \cos\theta}{\Delta f} - 1 \Bigr) N_p \nonumber\\
        &=
\begin{cases}
\bigl( \frac{M_p}{N_p} \cos\theta - 1 \bigr) N_p
   = M_p \cos\theta - N_p, & \text{Walker-}\delta \\[8pt]
\bigl( \frac{M_p}{2N_p} \cos\theta - 1 \bigr) N_p
   = \frac{M_p \cos\theta}{2} - N_p, & \text{Walker-star}
\end{cases}
\end{align}
When $\Delta f / \cos\theta > \Delta \Omega$, the intra-plane satellite spacing is larger, and the rectangular geometric relationship in \fig\ref{fig:l2adj}~(a) yields:
\begin{eqnarray}
\dfrac{\Delta f}{f_0^\star} = \dfrac{\Delta \Omega + f_0^\star / \cos\theta}{f_0^\star / \cos\theta}
\end{eqnarray}
and the reconfigured phase factor is directly obtained as:
\begin{align}
\label{eqt:l2f2}
F^\star &= \dfrac{f_0^\star}{\Delta f} N_p
        = \Bigl( 1 - \dfrac{\Delta \Omega \cos\theta}{\Delta f} \Bigr) N_p \nonumber\\
        &=
\begin{cases}
N_p - M_p \cos\theta, & \text{Walker-}\delta \\[8pt]
N_p - \dfrac{M_p \cos\theta}{2}, & \text{Walker-star}
\end{cases}
\end{align}
Since $\left( \frac{\Delta \Omega \cos\theta }{\Delta f} -1 \right) > 0$ when~$\Delta f / \cos\theta < \Delta \Omega$, and negative otherwise, combining \eqt\ref{eqt:l2f1} and \eqt\ref{eqt:l2f2} yields the final phase factor adjustment formula:
\begin{eqnarray}
F^\star = \left\{
\begin{aligned}
\left\lvert N_p - M_p \cos\theta\right\rvert  & , \quad \text{Walker-}\delta\\
\left\lvert N_p - \dfrac{M_p \cos\theta}{2}\right\rvert & ,\quad \text{Walker-star}	
\end{aligned}
\right.
\end{eqnarray}

\subsubsection{Reconfiguring to the~$\mathcal{L}_3$ layout}

\begin{figure}[htbp]
\centering
\includegraphics[width=0.5\linewidth]{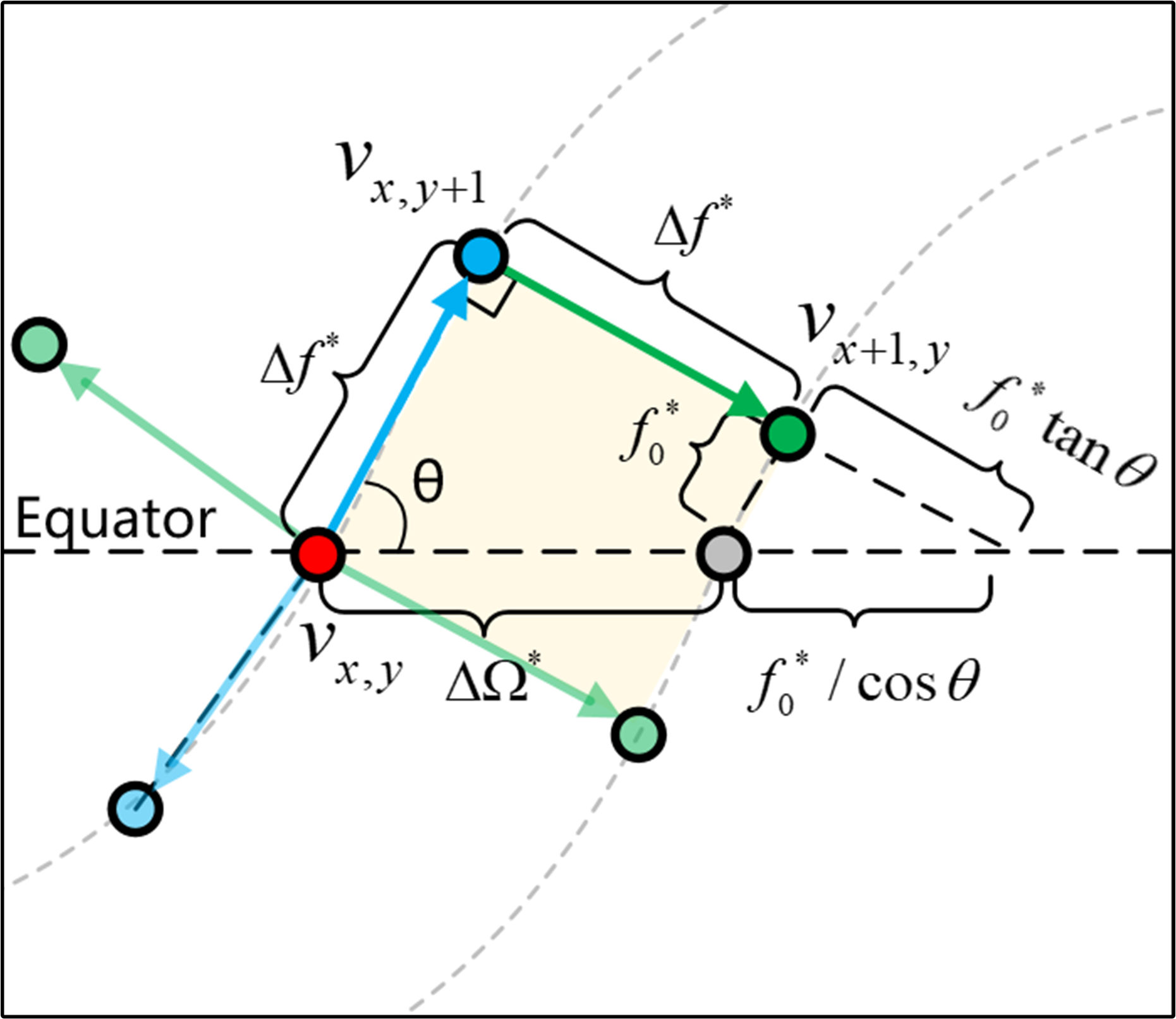}
\caption{Relative geometric configuration of satellites in the~$\mathcal{L}_3$~layout}
\label{fig:l3adj}
\end{figure}

To achieve the $\mathcal{L}_3$ layout, the $F$, $N_p$, and $M_p$ must all be jointly reconfigured, so that~$v_{x,y}$, $v_{x,y-1}$, $v_{x+1,y}$, and~$v_{x+1,y+1}$ form an approximately square configuration near the equator, as shown in \fig\ref{fig:l3adj}. In this case, the reconfigured phase difference~$\Delta f^\star$ is approximately equal to the RAAN spacing~$\Delta \Omega^\star$, and the reconfigured constellation satisfies the following geometric constraints:
\begin{gather}
\dfrac{f_0^\star \tan\theta}{\Delta f^\star} = \dfrac{f_0^\star / \cos\theta}{\Delta \Omega^\star}\\
\dfrac{\Delta f^\star + f_0^\star \tan\theta}{\Delta f^\star} = \tan\theta \\
N_p^\star M_p^\star = N_p M_p = N_M
\end{gather}
These constraints simplify to:
\begin{eqnarray}
\dfrac{\Delta \Omega^\star}{\Delta f^\star}&=& \dfrac{1}{\cos\theta \tan\theta} = 
\left\{
\begin{aligned}
\dfrac{M_p^\star}{N_p^\star}& , \quad \text{Walker-}\delta\\
\dfrac{M_p^\star}{2 N_p^\star}& ,\quad \text{Walker-star}
\end{aligned}
\right., \\  
\quad \dfrac{f_0^\star}{\Delta f^\star} &=& \dfrac{\tan\theta - 1}{\tan\theta}
\end{eqnarray}
Therefore, the reconfigured parameters are:
\begin{eqnarray}
N_p^\star &=& 
\left\{
\begin{aligned}
	 &\sqrt{N_M \cos\theta \tan\theta} , \quad \text{Walker-}\delta\\
	 &\sqrt{\dfrac{N_M \cos\theta \tan\theta}{2}} , \quad \text{Walker-star}
\end{aligned}
\right.\\
M_p^\star &=&  \dfrac{N_M}{N_p^\star} \\
F^\star &=& \dfrac{f_0 N_p^\star}{\Delta f} = \dfrac{(\tan\theta - 1) N_p^\star}{\tan\theta}
\end{eqnarray}

\subsubsection{Reconfiguring to the~$\mathcal{L}_4$ layout}

\begin{figure}[htbp]
\begin{center}
	\subfloat[$\Delta f/(2 \cos\theta) > \Delta \Omega$]{
		\includegraphics[width=0.48\linewidth]{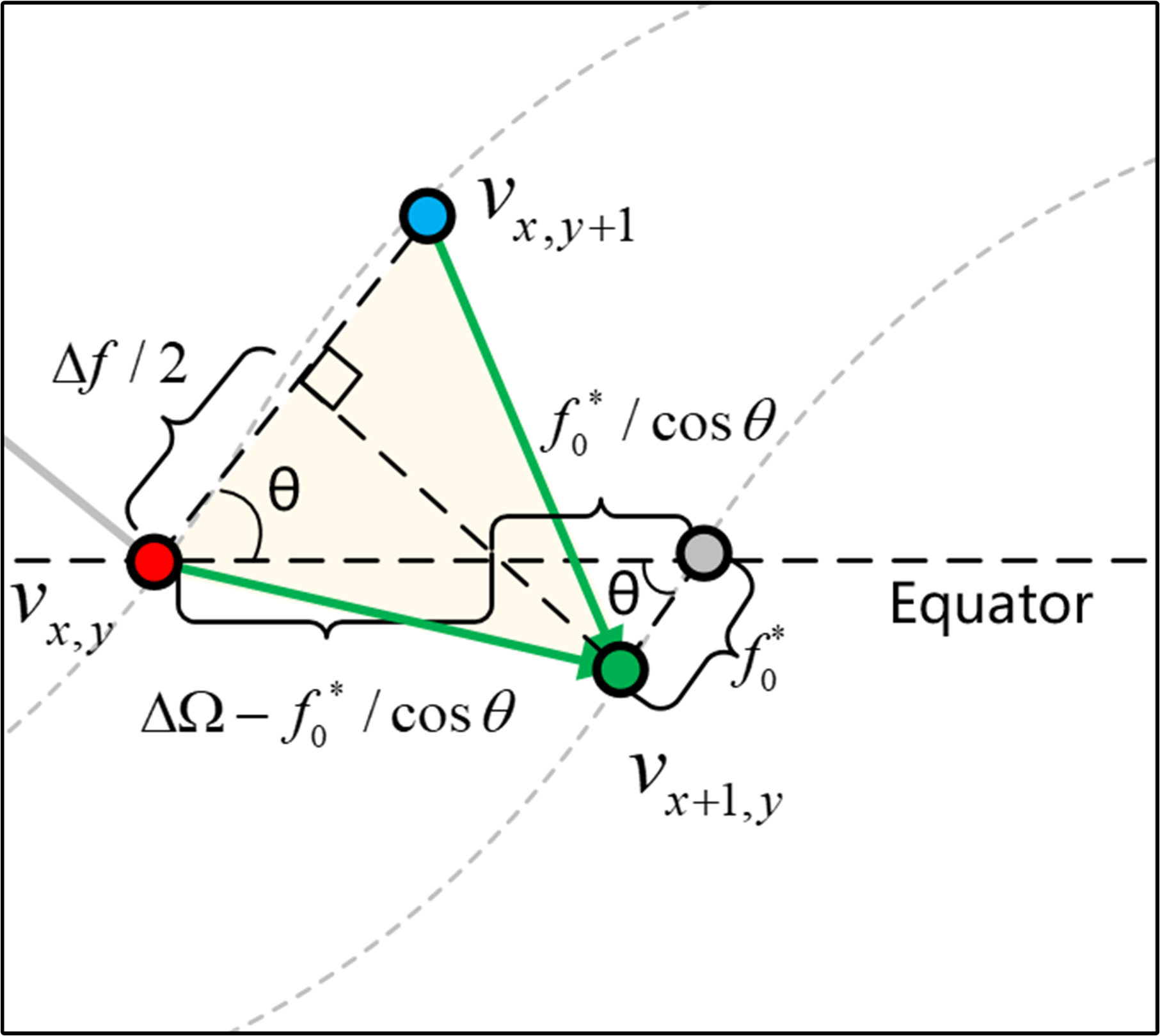}
	}\subfloat[$\Delta f/(2 \cos\theta) < \Delta \Omega$]{
		\includegraphics[width=0.48\linewidth]{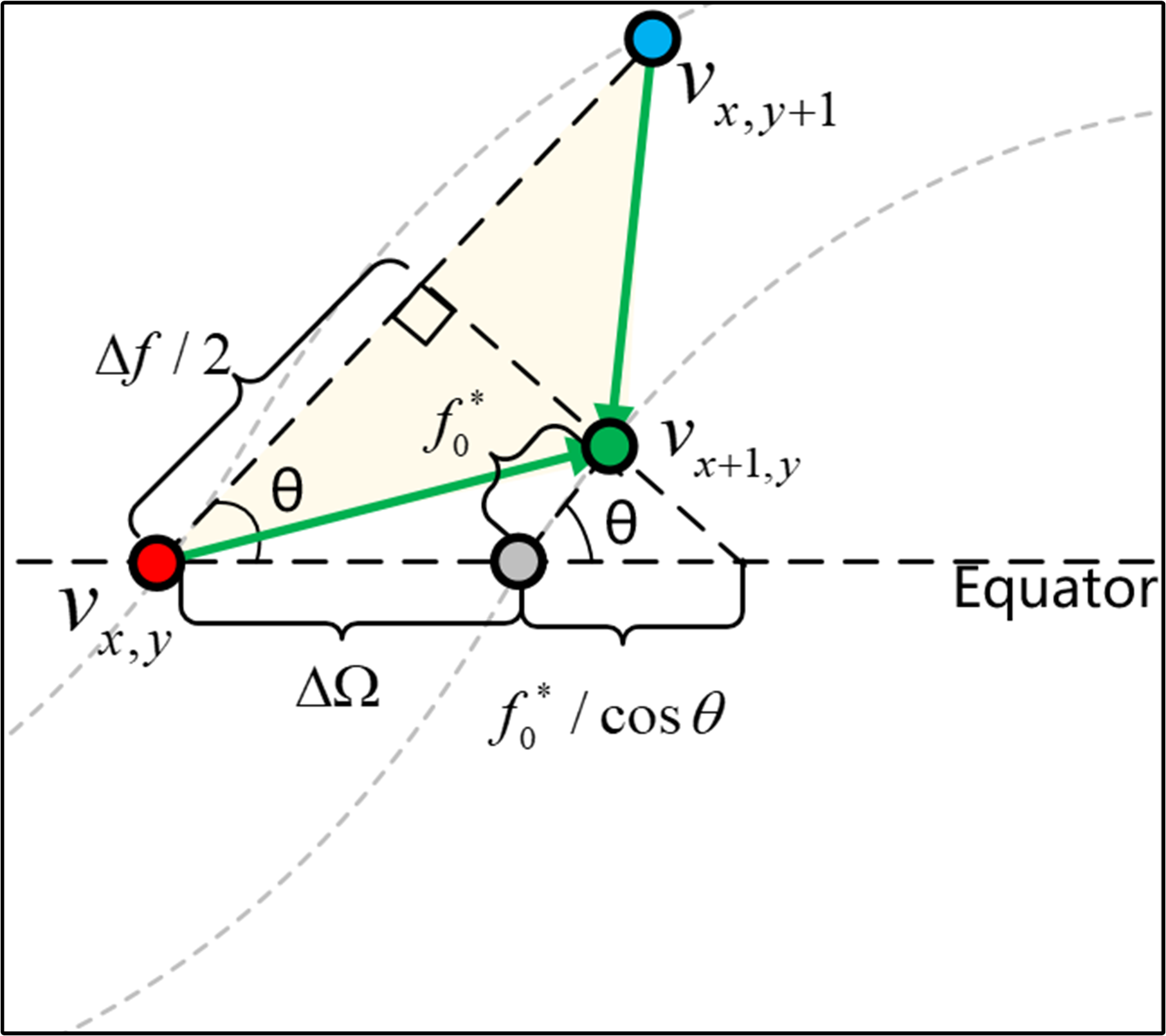}
	}
\caption{Relative geometric configuration of satellites in the~$\mathcal{L}_4$ layout}
\label{fig:l4adj}
\end{center}
\end{figure}

The~$\mathcal{L}_4$ layout is achieved solely by reconfiguring the phase factor~$F$, with the goal of making satellites~$v_{x,y}$, $v_{x,y-1}$, and~$v_{x+1,y}$ form an approximately isosceles triangle near the equator. Depending on the relative magnitudes of the RAAN spacing~$\Delta \Omega$ and the phase difference~$\Delta f$, two geometric cases can be distinguished, as shown in \fig\ref{fig:l4adj}.

When $\Delta f / (2 \cos\theta) > \Delta \Omega$, the intra-plane satellite spacing is larger, and the isosceles condition in \fig\ref{fig:l4adj}~(a) yields:
\begin{eqnarray}
\dfrac{f_0^\star}{\Delta f / 2} = \dfrac{f_0^\star / \cos\theta}{\Delta \Omega + f_0^\star / \cos\theta}
\end{eqnarray}
From the isosceles condition, the phase factor is obtained as:
\begin{eqnarray}
	\label{eqt:l4f1}
&F^\star &=  \dfrac{f_0^\star N_p}{\Delta f}  = (\dfrac{ \Delta \Omega}{\Delta f}\cos\theta -\dfrac{1}{2})N_p\\ & = &
\left\{
\begin{aligned}
 M_p \cos\theta - \dfrac{N_p}{2} & , \quad \text{Walker-}\delta\\
\dfrac{M_p \cos\theta -N_p}{2} & ,\quad \text{Walker-star}
\end{aligned}
\right.
\end{eqnarray}
When $\Delta f / (2 \cos\theta) < \Delta \Omega$, the inter-plane satellite spacing is larger, and the isosceles condition in \fig\ref{fig:l4adj}~(b) becomes:
\begin{eqnarray}
\dfrac{f_0^\star}{\Delta f / 2} = \dfrac{f_0^\star / \cos\theta}{\Delta \Omega - f_0^\star / \cos\theta}
\end{eqnarray}
Accordingly, the phase factor is obtained as:
\begin{align}
	\label{eqt:l4f2}
F^\star &=  \dfrac{f_0^\star N_p}{\Delta f}  = \left( \dfrac{1}{2} - \dfrac{\Delta \Omega}{\Delta f}\cos\theta \right) N_p \\
&= 
\left\{
\begin{aligned}
 \dfrac{N_p}{2} - M_p \cos\theta  & , \quad \text{Walker-}\delta\\
\dfrac{N_p - M_p \cos\theta }{2} & ,\quad \text{Walker-star}
\end{aligned}
\right.
\end{align}
Since $\left( \frac{\Delta \Omega \cos\theta }{\Delta f} -\dfrac{1}{2} \right) > 0$ when~$\Delta f / (2 \cos\theta) > \Delta \Omega$, and negative otherwise, combining \eqt\ref{eqt:l4f1} and \eqt\ref{eqt:l4f2} yields the final phase factor adjustment formula:
\begin{eqnarray}
F^\star = \left\{
\begin{aligned}
\left\lvert \dfrac{N_p}{2} - M_p \cos\theta\right\rvert  & , \quad \text{Walker-}\delta\\
\left\lvert \dfrac{N_p - M_p \cos\theta}{2}\right\rvert  & ,\quad \text{Walker-star}	
\end{aligned}
\right.
\end{eqnarray}

\subsubsection{Reconfiguring to the~$\mathcal{L}_5$ layout}
\begin{figure}[htbp]
\centering
\includegraphics[width=0.5\linewidth]{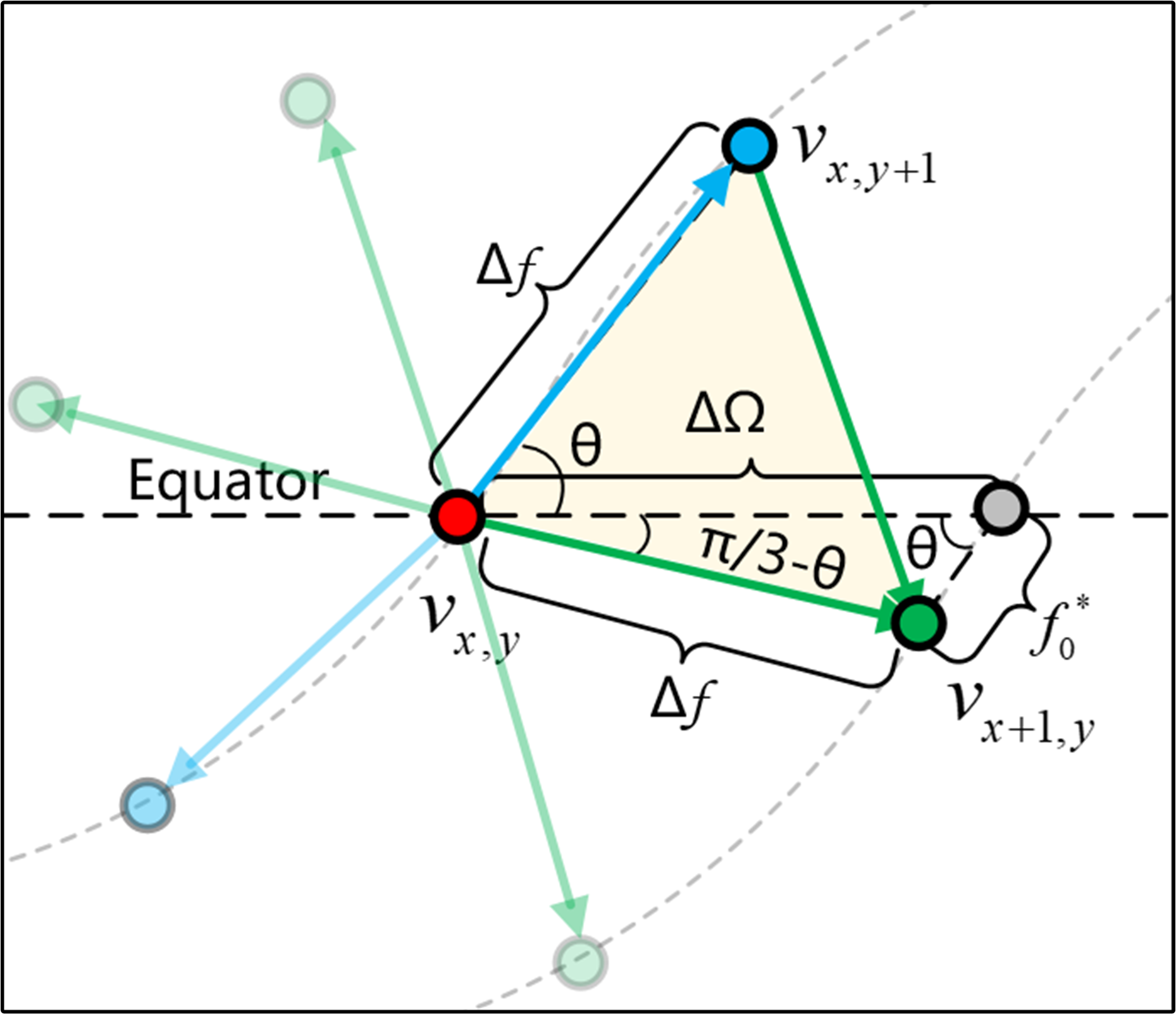}
\caption{Relative geometric configuration of satellites in the~$\mathcal{L}_5$ layout}
\label{fig:l5adj}
\end{figure}

To achieve the~$\mathcal{L}_5$ layout, the $F$, $N_p$, and $M_p$ must all be jointly reconfigured, so that~$v_{x,y}$, $v_{x,y-1}$, and~$v_{x+1,y}$ form an approximately equilateral triangle near the equator, as shown in \fig~\ref{fig:l5adj}.
Here, the intra-plane and inter-plane satellite distances are equal, and the apex angle of the triangle is approximately~$60^\circ$. The geometric constraints are:
\begin{eqnarray}
\dfrac{\Delta \Omega^\star}{\Delta f^\star} &=& \dfrac{\sqrt{3}}{2 \sin\theta} = 
\left\{
\begin{aligned}
\dfrac{M_p^\star}{N_p^\star}& , \quad \text{Walker-}\delta\\
\dfrac{M_p^\star}{2 N_p^\star}& ,\quad \text{Walker-star}
\end{aligned}
\right., \\  
\quad \dfrac{f_0^\star}{\Delta f^\star} &=& \dfrac{\sin (\pi/3 - \theta)}{\sin \theta}
\end{eqnarray}
The parameters are thus obtained as:
\begin{eqnarray}
N_p^\star &=&
\left\{
\begin{aligned}
\sqrt{\dfrac{2 N_M \sin \theta}{\sqrt{3}}} , \quad \text{Walker-}\delta\\
\sqrt{\dfrac{N_M \sin \theta}{\sqrt{3}}} ,\quad \text{Walker-star}
\end{aligned}
\right.
 \\
M_p^\star &=&  \dfrac{N_M}{N_p^\star}  \\
F^\star &=& \dfrac{f_0^\star N_p^\star}{\Delta f^\star} = \dfrac{\sin (\pi/3 - \theta) N_p^\star}{\sin \theta}
\end{eqnarray}

\subsection{Rationale for Designing a Uniform Constellation under Non-uniform, Dynamic Traffic Demands}
\label{app:traffic}

In practice, ground traffic demand is highly non-uniform in both space and time\cite{roughan2005simplifying}: the volumes carried by densely populated areas versus remote regions, daytime versus nighttime, and different latitude bands often differ by orders of magnitude, exhibiting strong spatiotemporal correlation. Existing studies widely acknowledge this discrepancy and mostly resort to more complex, finer-grained non-uniform traffic models, employing them as the basis for such tasks as routing optimization, network capacity planning, and constellation design\cite{wang2022intersatellite,bhattacherjee2019network,li2025small}. 
In contrast, this work adopts a uniform inter-satellite traffic model, i.e., it assumes that the traffic load is uniformly distributed among satellite nodes, which does not match the actual traffic characteristics. 
This raises the question: is a uniform, symmetric constellation design actually reasonable in the face of unbalanced ground demand?

We argue that, although the instantaneous loads of individual satellites are highly imbalanced in the short term owing to the spatiotemporal variability of ground demand, the high-speed motion of LEO satellites causes each satellite to traverse the same ground regions over a sufficiently long period. Consequently, the traffic carried by each satellite fluctuates periodically over time, and its long-term average load tends to become consistent.

\subsubsection{System Model}

Consider a LEO satellite constellation comprising~$N$~satellites, all with the same orbital altitude, inclination, and coverage capability. Let~$\mathcal{V} = \{1, 2, \ldots, N\}$ denote the set of satellites, where $\mathcal{V}$ is the vertex set of the network. 
Let the traffic demand density at ground location $\mathbf{x}$ at time $t$ be $\lambda(\mathbf{x}, t)$, with the Earth's surface $\mathcal{D}$ as its domain. Denote the ground coverage region of satellite $v$ at time $t$ by $\mathcal{A}_v(t) \subset \mathcal{D}$; then the instantaneous load of satellite $v$ is:

\begin{equation}
L_v(t) = \int_{\mathbf{x} \in \mathcal{A}_v(t)} \lambda(\mathbf{x}, t) \, d\mathbf{x}, \quad v = 1, 2, \ldots, N
\label{eq:instant_load}
\end{equation}

\subsubsection{Basic Assumptions}

\paragraph{Assumption 1 (Orbital periodicity):} In the absence of a repeat ground track orbit, the orbital motion of a LEO satellite is periodic over the long term. Let the orbital period be $T_0$; the ground track of satellite $v$ then satisfies:
\begin{equation}
\mathbf{p}_v(t + T_0) = \mathbf{p}_v(t) + \Delta \mathbf{p}
\end{equation}
where $\Delta \mathbf{p}$ is an equal drift in the longitudinal direction. After a finite number of repeat cycles, the ground track covers the entire serviceable latitude band.

\paragraph{Assumption 2 (Long-term traffic demand stability):} The long-term average demand density at each ground location exists and is finite:

\begin{equation}
\bar{\lambda}(\mathbf{x}) = \lim_{T \to \infty} \frac{1}{T} \int_0^T \lambda(\mathbf{x}, t) \, dt < \infty, \quad \forall \mathbf{x} \in \mathcal{D}
\label{eq:long_term_demand}
\end{equation}

\subsubsection{Lemma (Consistency of the Long-term Average Load):} For any satellite $v \in \mathcal{V}$, its long-term average load converges to the same constant, i.e., there exists $\bar{L} \in \mathbb{R}^+$ such that:

\begin{equation}
\lim_{T \to \infty} \frac{1}{T} \int_0^T L_v(t) \, dt = \bar{L}, \quad \forall v \in \mathcal{V}
\label{eq:convergence}
\end{equation}

Here:

\begin{equation}
\bar{L} = \frac{1}{|\mathcal{D}|} \int_{\mathcal{D}} \bar{\lambda}(\mathbf{x}) \, d\mathbf{x}
\label{eq:average_load}
\end{equation}

where $|\mathcal{D}|$ denotes the area of the serviceable region.

\subsubsection{Proof}

Substituting \eqt\ref{eq:instant_load} into the left-hand side of \eqt\ref{eq:convergence}:
\begin{equation}
\lim_{T \to \infty} \frac{1}{T} \int_0^T L_v(t) \, dt
= \lim_{T \to \infty} \frac{1}{T} \int_0^T \int_{\mathcal{A}_v(t)} \lambda(\mathbf{x}, t) \, d\mathbf{x} \, dt
\label{eq:proof_start}
\end{equation}

Interchanging the order of integration (by Tonelli's theorem, the order of integration can be interchanged for non-negative measurable functions):

\begin{equation}
= \lim_{T \to \infty} \int_{\mathcal{D}} \left[ \frac{1}{T} \int_0^T \lambda(\mathbf{x}, t) \cdot \mathbf{1}_{\{\mathbf{x} \in \mathcal{A}_v(t)\}} \, dt \right] d\mathbf{x}
\label{eq:fubini}
\end{equation}

where $\mathbf{1}_{\{\cdot\}}$ is the indicator function, which equals 1 when $\mathbf{x} \in \mathcal{A}_v(t)$ and 0 otherwise.

Define:
\begin{equation}
f_v(\mathbf{x}) = \lim_{T \to \infty} \frac{1}{T} \int_0^T \mathbf{1}_{\{\mathbf{x} \in \mathcal{A}_v(t)\}} \, dt
\label{eq:coverage_prob}
\end{equation}
here, $f_v(\mathbf{x})$ denotes the fraction of time that satellite $v$ covers location $\mathbf{x}$ over the long-term time scale.
Since the satellite orbits are periodic and the constellation is uniformly configured, the ground tracks of all satellites are equivalent in the long-term statistical sense. Over a sufficiently long time window, the coverage time of each satellite over every global latitude band converges to the same value. Therefore:
\begin{equation}
f_v(\mathbf{x}) = f(\mathbf{x}), \quad \forall v \in \mathcal{V}, \quad \int_{\mathcal{D}} f(\mathbf{x}) \, d\mathbf{x} = 1
\label{eq:coverage_equal}
\end{equation}
Substituting \eqt\ref{eq:coverage_prob} and \eqt\ref{eq:coverage_equal} into \eqt\ref{eq:fubini} and applying the dominated convergence theorem:
\begin{align}
\lim_{T \to \infty} \frac{1}{T} \int_0^T L_v(t) \, dt
&= \int_{\mathcal{D}} \bar{\lambda}(\mathbf{x}) \cdot f(\mathbf{x}) \, d\mathbf{x} \notag \\
&= \frac{1}{|\mathcal{D}|} \int_{\mathcal{D}} \bar{\lambda}(\mathbf{x}) \, d\mathbf{x} \notag \\
&= \bar{L}
\label{eq:final}
\end{align}
where the last step exploits the uniformity of the coverage-time fraction $f(\mathbf{x}) = 1/|\mathcal{D}|$.
The right-hand side of \eqt\ref{eq:final} is independent of the satellite index $v$, and therefore holds for all satellites:

\begin{equation}
\lim_{T \to \infty} \frac{1}{T} \int_0^T L_v(t) \, dt = \bar{L}, \quad \forall v \in \mathcal{V}
\label{eq:conclusion}
\end{equation}
The lemma is thus proved.

The above mathematical result reveals an important system property: although the ground traffic demand is highly non-uniform in both space and time, the high-speed motion of LEO satellites enables each satellite to traverse the same spatial regions over a sufficiently long period. Consequently, although the instantaneous load of a single satellite fluctuates sharply, its long-term statistical characteristics tend to become consistent. This provides a theoretical basis for the rationality of a uniform, symmetric constellation design.

\bibliographystyle{IEEEtran}
\bibliography{ref.bib}
\newpage

\end{document}